\documentclass[fleqn,usenatbib]{mnras}
\pdfoutput=1
\usepackage[T1]{fontenc}
\usepackage[latin9]{inputenc}
\usepackage{graphicx, bm, color, babel}
\usepackage{hyperref}
\usepackage{amsmath,amssymb}
\usepackage{mathtools}
\usepackage{xcolor}
\usepackage{tikz}
\usepackage{hyperref}
\usepackage{natbib}

\usetikzlibrary{arrows.meta,positioning,calc}
\def\be{\begin{equation}}
\def\ee{\end{equation}}
\def\bea{\begin{eqnarray}}
\def\eea{\end{eqnarray}}

\renewcommand{\subsubsection}[1]{~\\[-3mm]
\noindent{\textbf{\emph{#1}}} }

\renewcommand{\k}{{\bm k}}
\newcommand{\x}{{\bm x}}
\newcommand{\n}{{\bm n}}
\newcommand{\q}{{\bm q}}
\newcommand{\s}{{\bm s}}
\newcommand{\y}{{\bm y}}

\newcommand{\vvel}{{\bm v}}
\renewcommand{\d}{{\bm d}} % Pair center

\newcommand{\<}{\langle}
\renewcommand{\>}{\rangle}

\newcommand{\diff}{\mathrm{d}}

\newcommand{\Pobs}{P^{\rm(O)}}          % The Observer Spectrum
\newcommand{\barPobs}{{\bar P}^{\rm(O)}}          % The Observer Spectrum
\newcommand{\Bobs}{B^{\rm(O)}}          % The Observer Bispectrum
\newcommand{\Pkd}{\bar{P}}              % The Mixed k-d spectrum
\newcommand{\Plc}{\tilde{P}}            % The Non-diagonal lightcone kernel
\newcommand{\K}{\mathcal{K}}            % The Kernel (Kaiser/GR)
\newcommand{\LC}{{\rm LC}}

\newcommand{\pritha}[1]{{\color{black}{#1}}}

\title[Observer power spectrum]{The observer power spectrum  for lightcone statistics, \\ integrated relativistic observables and wide angle effects
}

\author[Clarkson and Paul]{
Chris Clarkson,$^{1}$\thanks{E-mail: chris.clarkson@qmul.ac.uk}
Pritha Paul$^{1,2}$\thanks{E-mail: pritha.paul@physik.uni-muenchen.de}
\\
$^{1}$Centre for Theoretical Physics and Astronomy, Queen Mary University of London, Mile End Road, London E1 4NS, United Kingdom\\
$^{2}$University Observatory, Faculty of Physics, Ludwig-Maximilians-Universit\"at, Scheinerstr. 1, 81679 Munich, Germany
}

\date{Accepted XXX. Received YYY; in original form ZZZ}

\pubyear{2026}

\begin{document}

%\title{The observer power spectrum  for lightcone statistics, \\ integrated relativistic observables and wide angle effects}

%\author{Chris Clarkson}
%\email{chris.clarkson@qmul.ac.uk}
%\affiliation{Centre for Theoretical Physics and Astronomy, Queen Mary University of London, Mile End Road, London, E1 4NS, United Kingdom}

%\author{Pritha Paul}
%\email{pritha.paul@physik.uni-muenchen.de}
%\affiliation{Centre for Theoretical Physics and Astronomy, Queen Mary University of London, Mile End Road, London, E1 4NS, United Kingdom}
%\affiliation{University Observatory, Faculty of Physics,  Ludwig-Maximilians-Universit\"at, 
%Scheinerstr.~1, 81679 Munich, Germany}

\label{firstpage}
\pagerange{\pageref{firstpage}--\pageref{lastpage}}

\maketitle

\begin{abstract}

The statistics of large-scale structure are naturally described by power spectra in Fourier space. For fields on spatial hypersurfaces, translational invariance makes different Fourier modes uncorrelated and the power spectrum diagonal. Cosmological observables, however, are measured on our past lightcone, where wide-angle effects, radial evolution and integrated effects such as lensing break this symmetry: Fourier-space statistics become non-diagonal, with mode-mixing generated by the geometry of the lightcone itself. We define a more natural observer power spectrum by Fourier transforming over observer positions on a spatial hypersurface with fixed lightcone coordinates, rather than over source positions on a single lightcone. This forms a field on the observer hypersurface with a moveable light-ray leg. Statistical homogeneity of the observer hypersurface implies that this spectrum is diagonal for any observable, whether local or integrated and does not suffer from mode-mixing. 
%For local fields the observer spectrum is just the unequal-time spatial power spectrum with a geometric phase; for integrated quantities such as lensing convergence it factorises into line-of-sight transfer functions acting on each leg independently, with no mode-mixing. 
We show how the various two-point statistics used in large-scale structure analysis
%~-- the unequal-time power spectrum, the mixed-space spectrum, the non-diagonal lightcone power spectrum, angular power spectra, and the Yamamoto estimator~-- 
are each recovered as projections of the observer spectrum.
%, clarifying the origin of wide-angle corrections and non-diagonality at each level. 
This extends directly to higher-order statistics. We illustrate it by constructing the relativistic kernel for the observed galaxy number count fluctuation, including density, redshift-space distortions, Doppler, lensing magnification, and integrated Sachs-Wolfe contributions.\\

\end{abstract}

\maketitle

\section{Introduction}
\label{sec:intro}

The power spectrum of a cosmological field on a spatial hypersurface at
fixed time is diagonal, in the sense that statistical homogeneity makes
Fourier modes with different wavevectors uncorrelated, so all two-point
information is contained in a single function $P(k)$. This follows from translational invariance and does not require the field to be linear or Gaussian. For galaxy clustering in redshift space
this acquires an angular dependence through the Kaiser effect
\citep{1987MNRAS.227....1K}, but the plane parallel equal redshift spectrum remains diagonal and forms
the basis of the standard analysis pipeline from FKP
\citep{1994ApJ...426...23F} to Yamamoto-type estimators
\citep{2006PASJ...58...93Y,2015MNRAS.453L..11B,2015PhRvD..92h3532S}.
However, observables in galaxy surveys are not measured on spatial
hypersurfaces, but on the observer's past lightcone which complicates this simple picture. Radial
evolution means that the growth factor, bias and selection function vary
with distance; wide-angle geometry means that galaxies in a pair have
different lines of sight; and integrated effects such as lensing
magnification depend on the gravitational potential along the photon
path, not just at the source. If one Fourier transforms such a lightcone
field over source positions, the transform no longer follows the
translation symmetry of a single spatial slice: radial evolution,
wide-angle geometry and integrated terms generate non-diagonal
correlations between Fourier modes
\citep{1999ApJ...517....1Y,1999ApJ...527..488Y,2013JCAP...11..044D,
2018JCAP...03..019T,2022JCAP...01..061C}. This is why several different
two-point statistics are used in the literature, each adapted to a
different projection of the same lightcone correlations.

For galaxy correlations, the most common Fourier-space approach is to transform the lightcone 2-point correlation function over the pair separation~$\bm{s}$ at fixed midpoint or endpoint~$\d$, giving a mixed spectrum~$\bar{P}(\k,\d)$ that depends on both a Fourier mode and a configuration-space position. The Kaiser effect implies this is an even polynomial in $\mu=\hat\k\cdot\hat\d$ in the plane parallel limit.  In multi-tracer or cross-correlations, where the two populations have different bias, magnification bias or evolution bias, a Doppler contribution generates odd multipoles that provide a route to detecting general-relativistic effects~\citep{2014PhRvD..89h3535B,2017JCAP...01..032G,2021JCAP...12..003F}. Wide-angle corrections to $\bar{P}$ are developed as perturbative expansions in $s/d$~\citep{1998ApJ...498L...1S,2000ApJ...535....1M,2004ApJ...614...51S,2008MNRAS.389..292P,2016JCAP...01..048R,2018MNRAS.476.4403C,2019JCAP...03..040B,2023JCAP...04..067P}, and including relativistic Doppler corrections within this framework shows that the leading wide-angle and leading relativistic contributions enter at the same order and must be treated jointly~\citep{2012JCAP...10..025B,2015MNRAS.447.1789Y,2023JCAP...04..067P,2024JCAP...08..027J}.  
Alternatively, one can Fourier transform the lightcone field over source positions directly, giving a non-diagonal spectrum $\tilde{P}(\k_1,\k_2)$ in which radial evolution and wide-angle geometry correlate different Fourier modes~\citep{1999ApJ...517....1Y,1999ApJ...527..488Y}. 
In a spherical harmonic basis, a different partial transform is obtained by keeping the radial positions in configuration space and expanding only the angular dependence at the observer. The resulting angular power spectrum between two redshift shells, $C_\ell(z_1,z_2)$, naturally respects rotational invariance and treats local and integrated contributions on the same footing~\citep{1994MNRAS.266..219F,1995MNRAS.275..483H,1998ASSL..231..185H,2013JCAP...11..044D,2013PhRvD..88b3502Y,2018JCAP...03..019T,2018JCAP...10..032T}. One may also transform the radial dependence with spherical Bessel functions, giving a fully three-dimensional harmonic description in which angular and radial modes are separated~\citep{1994MNRAS.266..219F,1995MNRAS.275..483H,2018MNRAS.476.4403C,2024PhRvD.109h3502G,2024arXiv240404812W,2024arXiv240404811B}. These approaches are especially well suited to full-sky relativistic and wide-angle calculations, but they are not the same as a Cartesian Fourier transform over a single lightcone.
%: the diagonal variable is angular momentum, or angular momentum together with a radial wavenumber, rather than the wavevector used in standard Fourier-space estimators.

These issues are particularly important for the fully relativistic observed number-count fluctuation. Its linear expression contains local density and Kaiser RSD terms, Doppler and potential contributions, lensing magnification, and other integrated effects along the photon path~\citep{2009PhRvD..80h3514Y,2010PhRvD..82h3508Y,2011PhRvD..84f3505B,2011PhRvD..84d3516C,2012PhRvD..85b3504J,2013JCAP...11..044D,2021JCAP...12..009M}. These terms have been incorporated into full-sky angular spectra and correlation functions~\citep{2013JCAP...11..044D,2018JCAP...03..019T,2018JCAP...10..032T}, where the full contribution of these effects is included. They are also included into wide-angle Fourier-space power spectra~\citep{2022JCAP...01..061C,2020JCAP...07..048B,2023PhRvL.131k1201F,2026JCAP...06..039A}, and into wide-separation expansions of bispectrum multipoles~\citep{2025JCAP...04..080A}, but with with necessary approximations due to the mismatch of the lightcone geometry and the plane-wave expansion. Closest to the viewpoint developed here is the theory power spectrum~\citep{2020JCAP...11..064G}, which defines a diagonal Fourier-space spectrum for the observed galaxy fluctuation at linear order. What remains less explicit is how these different statistics arise from the same lightcone correlations, and how to separate the intrinsic statistics of the observable from the lightcone geometry and the projection imposed by a particular estimator.

This motivates the main aim of this paper: to give a common Fourier-space description of lightcone observables, including integrated effects such as lensing, in which the role of statistical homogeneity or translational invariance remains explicit. We do this by starting from a single diagonal observer spectrum, whose definition is exact for the observable under consideration, and then deriving the usual lightcone statistics as integral transforms (or `projections') of it. This makes clear which transforms generate mode-mixing, where wide-angle geometry, radial evolution and line-of-sight terms enter, and which further approximations are made when those projections are expanded perturbatively.

In this paper we therefore define the observer power spectrum $\Pobs$, together with its higher-order generalisations, which keeps the diagonal Fourier-space structure implied by statistical homogeneity while still describing quantities observed on the lightcone. This allows us to separate the intrinsic statistics of the observable from the projection used to describe it, and makes translation invariance of the field explicit. Instead of Fourier transforming over source positions on a single lightcone, we transform over observer positions on a homogeneous spatial hypersurface, at fixed light-ray directions and distances. The observable is then a field on the observer hypersurface, with a light-ray `leg' attached to each observer. Since the observers share a common spatial slice, statistical homogeneity implies that $\Pobs$ is diagonal in $\k$ for any observable, local or integrated. The lightcone dependence is not removed, but is contained in the external labels $(\chi_i,\hat{\n}_i)$.

For a local field, $\Pobs$ is just a  phase times the unequal-time spatial power spectrum. For an integrated quantity such as lensing convergence, $\Pobs$ factorises into line-of-sight transfer functions acting independently on each leg, proportional to $P_\Phi(k)$, the intrinsic potential power spectrum, at the same $\k$. This contrasts with source-space lightcone Fourier spectra, where the line-of-sight projection appears as a convolution over Fourier modes. The usual lightcone correlation function, angular spectra, mixed-space spectra, non-diagonal source-space spectra and Yamamoto multipoles are recovered by integrating or projecting $\Pobs$ in different ways. Thus $\Pobs$ provides a common object from which the usual lightcone statistics follow, and makes explicit where wide-angle effects, radial evolution, integrated terms and estimator projections enter. The same construction extends directly to bispectra and to $N$-point functions. Section~\ref{sec:theory} defines the observer Fourier transform and derives its relation to the standard two-point statistics; Section~\ref{sec:application} illustrates $\Pobs$ for the relativistic galaxy number-count fluctuation; and the Appendix derives the corresponding wide-angle expansion.

\section{The Observer Power Spectrum}
\label{sec:theory}

\subsection{Fourier Transforms}

We assume a spatially flat FLRW background with conformal time $\eta$. An observer at comoving position $\x_0$ on the hypersurface at $\eta_0$ receives light from a source at radial comoving distance $\chi$ in direction $\hat{\n}$, located at spacetime position in the background\footnote{Here and throughout, $\chi$ and $\hat{\n}$ label points on the background lightcone. Observable quantities  such as redshift, angle and luminosity distance may be used instead, with the conversion $\chi=\chi(z)$ made only after choosing a background cosmology. In the explicit linear examples below, integrated terms are evaluated along the background ray at the observed source position, i.e. in the Born approximation; perturbations of the light path, including post-Born corrections, would enter as additional contributions to the observable kernel.}
\begin{equation}
  \x_s(\x_0; \chi, \hat{\n}) = \x_0 + \chi \hat{\n}, \qquad \eta_s = \eta_0 - \chi.
\end{equation}
Let $X(\x_0; \chi, \hat{\n})$ denote some observable measured by this observer at the event $(\eta_s, \x_s)$. This may be a purely local field such as the galaxy number count in the Kaiser approximation, an integrated quantity such as the lensing convergence, or the measured number density itself, which is a combination of both.  We define the Observer Fourier Transform (OFT) by transforming with respect to the observer position $\x_0$ for a fixed light-ray $(\chi, \hat{\n})$:
\begin{equation}
  X^{\rm(O)}(\k; \chi, \hat{\n}) \equiv \int \diff^3x_0 \, e^{-i\k\cdot\x_0} \, X(\x_0; \chi, \hat{\n}).
  \label{eq:OFT_def}
\end{equation}
The inverse transform recovers the field at a specific observer location:
\begin{equation}
  X(\x_0; \chi, \hat{\n}) = \int \frac{\diff^3 k}{(2\pi)^3} \, e^{i\k\cdot\x_0} \, X^{\rm(O)}(\k; \chi, \hat{\n}).
  \label{eq:OFT_inverse}
\end{equation}
In particular, setting $\x_0 = {0}$ gives the observable in real space for an observer at the origin as an integral of the OFT over Fourier space:
\begin{equation}
  X(\bm{0}; \chi, \hat{\n}) = \int \frac{\diff^3 k}{(2\pi)^3} \, X^{\rm(O)}(\k; \chi, \hat{\n}).
  \label{eq:OFT_at_origin}
\end{equation}

This OFT differs fundamentally from the standard Fourier transform, which integrates over source positions $\x_s$ (usually with $\x_0=0$): 
\be\label{ksdjnvskjbnvsk}
\tilde{X}(\k) = \int \diff^3 x_s\, e^{-i\k\cdot\x_s} X(\x_s).
\ee
We normally have the source positions fixed to a constant-time hypersurface, which is fine provided $X(\x_0; \chi, \hat{\n})=X(\x_s)$ depends only on quantities local to the source, such as the density or the projected source velocity (so it can depend on $\n$). In this case the integration domain is $\mathbb{R}^3$. The problem with \eqref{ksdjnvskjbnvsk} arises when we have lightcone variables involved which link source and observer. In this case the observable $X$ depends on the radial distance $\chi$ and direction $\n$ both through the time evolution of the field and potentially through line-of-sight integrals, as for lensing. The integration domain in \eqref{ksdjnvskjbnvsk} is then restricted to the lightcone rather than a spatial slice, and the integrand varies with $|\x_s|=\chi$ through the growth factor, bias evolution, and any integrated kernels. This breaks translational invariance: the resulting two-point function is generically non-diagonal~\citep{1999ApJ...517....1Y,1999ApJ...527..488Y},
\be
\<\tilde{X}(\k_1)\tilde{X}(\k_2)\> = \Plc(\k_1, \k_2) \neq (2\pi)^3\delta^{(3)}_{\rm D}(\k_1+\k_2)\,P(k),
\ee
with different Fourier modes correlated by the radial structure of the lightcone. The non-diagonality is not a consequence of inhomogeneity in the underlying fields~-- the universe remains statistically homogeneous~-- but of the fact that the past lightcone is defined relative to a single observer at its apex, breaking the translational symmetry that a spatial hypersurface possesses.

We can think of $X(\x_0; \chi, \hat{\n})$ as a field on the observer hypersurface with a `light-ray leg' attached to it, which samples a single null ray in the spacetime between the source and the observer. The OFT does not have to deal with Fourier transforms over lightcones because we integrate over $\x_0$ instead of $\x_s$. Since all observers lie on the same spatial hypersurface at $\eta_0$, the integration domain is always a full spatial slice regardless of whether $X$ involves local or integrated quantities. Translational invariance is therefore preserved by construction, as we now show.

\begin{figure*}[t]
\centering
\includegraphics[width=\textwidth]{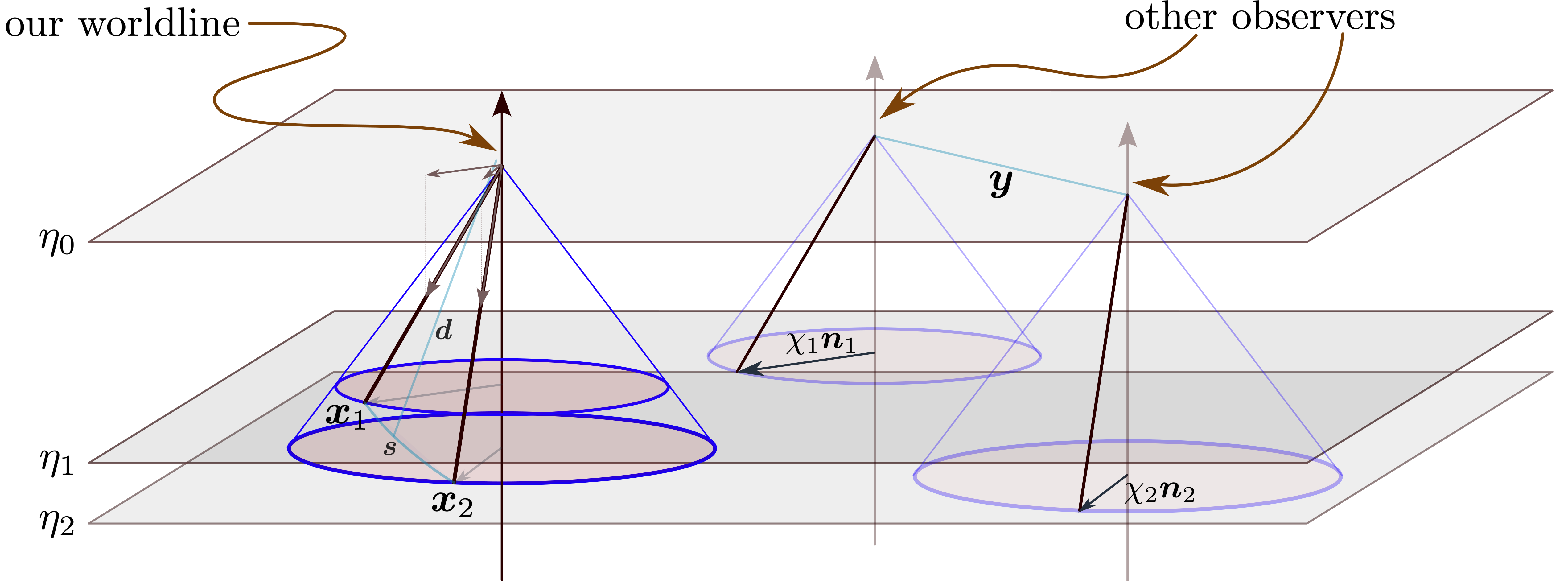}
\caption{\emph{Left:}
a single observer at $\bm{x}_0$ on the spatial hypersurface $\eta_0$
receives light from two sources at positions
$\bm{x}_1 = \bm{x}_0 + \chi_1\hat{\bm{n}}_1$ and
$\bm{x}_2 = \bm{x}_0 + \chi_2\hat{\bm{n}}_2$ on their 
past lightcone, intersecting the hypersurfaces $\eta_1$ and $\eta_2$.
Each null ray
samples the spacetime between observer and source; for integrated
observables such as lensing, the signal depends on the gravitational
potential along the entire ray, not just at the source. The pair separation $\bm{s} = \bm{x}_1 - \bm{x}_2$ and midpoint
$\bm{d}$ are the variables used in standard Fourier-space analyses
such as the mixed spectrum $\bar{P}(\bm{k},\bm{d})$ which FT over the lightcone. 
\emph{Right:} two further observers on the same hypersurface,
separated by $\bm{y}$, each looking along the same lightcone
directions $(\chi_i, \hat{\bm{n}}_i)$.  The OFT
integrates over this ensemble of observer positions with fixed legs
$(\chi_i, \hat{\bm{n}}_i)$. Because the integration domain is
always a spatial slice, statistical homogeneity guarantees that the
observer spectrum $P^{\rm(O)}(\bm{k})$ is diagonal~-- different
Fourier modes are uncorrelated regardless of whether the observable
is local or integrated along the null rays. The two-observer
correlation function $\Xi(\bm{y})$ is the configuration-space
FT of $P^{\rm(O)}(\bm{k})$, and the standard single-observer
lightcone correlation $\xi_{\rm LC}$ is recovered in the limit
$\bm{y} = 0$.}
\label{fig:lightcones}
\end{figure*}

\subsection{The Observer Spectrum}
\label{sec:observer_spectrum_def}

Consider the correlation of OFT modes:
\begin{equation}
  \big\< X^{\rm(O)}(\k_1; \chi_1, \hat{\n}_1)\,
        X^{\rm(O)}(\k_2; \chi_2, \hat{\n}_2) \big\>.
\end{equation}
Since the OFT integrates over the observer position $\x_0$ on a spatial hypersurface, statistical homogeneity guarantees that modes with different wavevectors are uncorrelated. Note that the legs $(\chi_i, \hat{\n}_i)$ are held fixed in the ensemble average. This allows us to define the Observer Power Spectrum $\Pobs$ via
\begin{align}
  &\big\< X^{\rm(O)}(\k_1; \chi_1, \hat{\n}_1)\,
        X^{\rm(O)}(\k_2; \chi_2, \hat{\n}_2) \big\> \nonumber\\ 
  &= (2\pi)^3 \delta_{\rm D}^{(3)}(\k_1 + \k_2)\,
    \Pobs(\k_1; \chi_1, \chi_2, \hat{\n}_1, \hat{\n}_2). 
  \label{eq:Pobs_def}
\end{align}
Consequently $\Pobs$ is diagonal in $\k$, just like the equal-time power spectrum on a spatial slice for a local quantity. To see this explicitly, write
$\< X^{\rm(O)}(\k_1;1)\,X^{\rm(O)}(\k_2;2)\>
= \int \diff^3x_0\,\diff^3x_0'\;
  e^{-i(\k_1\cdot\x_0+\k_2\cdot\x_0')}\,
  \<X(\x_0;1)\,X(\x_0';2)\>$.
Statistical homogeneity implies the configuration space two point correlation function (2pcf) is
$\<X(\x_0;1)\,X(\x_0';2)\>=\Xi(\x_0'-\x_0;\,1,2)$.
Setting $\y=\x_0'-\x_0$, the phase becomes
$(\k_1+\k_2)\cdot\x_0+\k_2\cdot\y$, so the $\x_0$ integral
gives $(2\pi)^3\delta^{(3)}_{\rm D}(\k_1+\k_2)$ and the $\y$
integral gives $\Pobs$. The difference is that instead of depending only on $(k, \mu)$ relative to a single line of sight, $\Pobs$ carries the full lightcone geometry through the radial distances $\chi_1, \chi_2$ and sky directions $\hat{\n}_1, \hat{\n}_2$. The wavevector $\k$ encodes the Fourier scale and direction of the underlying fluctuations, while the lightcone coordinates $(\chi_i, \hat{\n}_i)$ encode how those fluctuations are sampled along the two null rays. All wide-angle, relativistic and line-of-sight effects enter through this external dependence rather than through mode-mixing between different $\k$.

\subsubsection{The Local Field Limit}
\label{sec:local_field_limit}

For a local observable~-- one that depends only on quantities at the source position, such as the density or projected velocity~-- the OFT takes a simple form. Since $(\chi, \hat{\n})$ are held fixed, the displacement $\chi\hat{\n}$ between observer and source is a constant translation. Changing variables to the source position $\y = \x_0 + \chi\hat{\n}$, the transform becomes
\begin{align}
  X^{\rm(O)}(\k; \chi, \hat{\n}) 
  &= e^{i k\mu \chi} \int \diff^3y \, e^{-i\k\cdot\y} \, X(\y, \eta_0{-}\chi; \hat{\n}) \nonumber\\ 
  &= e^{i k\mu \chi} \, X(\k, \eta_0{-}\chi; \hat{\n}), 
\end{align}
where $\mu = \hat{\k}\cdot\hat{\n}$ and $X(\k; \eta, \hat{\n})$ is the standard spatial Fourier transform on the hypersurface at time $\eta$. The OFT factorises into the lightcone geometry encoded in the phase $e^{ik\mu\chi}$, which accounts for the spatial offset between observer and source, while the local physics is contained in $X(\k, \eta; \hat{\n})$, the normal FT.

Substituting into \eqref{eq:Pobs_def} and using 
$\< X(\k_1, \eta_1; \hat{\n}_1)\, X(\k_2, \eta_2; \hat{\n}_2) \> 
= (2\pi)^3 \delta_{\rm D}^{(3)}(\k_1 + \k_2) \, P(\k_1; \eta_1, \eta_2, \hat{\n}_1, \hat{\n}_2)$, 
we obtain
\begin{equation}
  \Pobs_{\rm local}(\k; \chi_1, \chi_2, \hat{\n}_1, \hat{\n}_2) 
  = e^{ik(\mu_1 \chi_1 - \mu_2 \chi_2)} \, P(\k; \eta_1, \eta_2, \hat{\n}_1, \hat{\n}_2),
  \label{eq:Pobs_local_result}
\end{equation}
where $\mu_i = \hat{\k}\cdot\hat{\n}_i$ and $\eta_i = \eta_0 - \chi_i$. Here and below $P$ is the unequal-time power spectrum of the underlying field on spatial slices at $\eta_1$ and $\eta_2$, with the lines of sight $\hat{\n}_i$ entering as field parameters. This relation is exact and holds in the non-linear regime. Note that $P$ is generically complex when the kernel includes
terms odd in $\mu$ such as the Doppler contribution $\propto
i\mu$. In this case $P$ is Hermitian:
$P(\k;\eta_1,\eta_2,\hat{\n}_1,\hat{\n}_2) =
P^*(\k;\eta_2,\eta_1,\hat{\n}_2,\hat{\n}_1)$, with the
imaginary part antisymmetric under $1\!\leftrightarrow\!2$.

The observer power spectrum for local fields is therefore the product of a geometric phase factor $e^{i\k\cdot(\chi_1\hat{\n}_1 - \chi_2\hat{\n}_2)}$ and the underlying spatial field statistics $P$. Wide-separation effects are included in the combination of the phase and the angular dependence intrinsic in $P$. In the plane-parallel limit ($\chi_1 \approx \chi_2 \equiv \bar\chi$, $\hat{\n}_1 \approx \hat{\n}_2 \equiv \hat{\n}$), the phase drops out, which gives the standard spatial hypersurface power spectrum $P(k, \mu; z(\bar\chi))$. In the plane-parallel limit ($1 = 2$), the Hermitian symmetry
reduces to $P = P^*$ for a single tracer. Odd-parity parts
therefore vanish in this limit --- the odd multipoles $P_1$,
$P_3$ are purely wide-separation effects.

\subsubsection{The Integrated Case}
\label{sec:lensing_integrated}

For an integrated observable such as the lensing potential $\psi$, the signal is a weighted integral along the line of sight. For a source at distance $\chi$, the lensing potential is
\begin{equation}
    \psi(\x_0; \chi, \hat{\n})
    = \int_0^\chi \diff\chi'\, W_L(\chi, \chi')\,
      \Phi_W(\x_0 + \chi'\hat{\n}, \eta_0 - \chi'),
\end{equation}
where $W_L(\chi,\chi') = (\chi-\chi')/(\chi\,\chi')$ is the lensing kernel and $\Phi_W=\Phi+\Psi$ is the Weyl potential~\citep{2001PhR...340..291B,2011PhRvD..84d3516C}.  To transform this, the OFT acts only on the observer position $\x_0$ holding the line of sight leg fixed in length and direction. This means we can interchange the $\x_0$- and $\chi'$-integrals and perform, at each fixed $\chi'$, exactly the same change of variables as in the local case, $\y = \x_0 + \chi'\hat{\n}$. This gives
\begin{equation}\label{lensexp}
    \psi^{\rm(O)}(\k; \chi, \hat{\n})
    = \int_0^\chi \diff\chi'\, W_L(\chi,\chi')\,
      e^{i k \mu\,\chi'}\,
      \Phi_W(\k, \eta_0 - \chi'),
\end{equation}
with $\mu = \hat{\k}\cdot\hat{\n}$ and $\Phi_W(\k,\eta)$ the standard spatial Fourier transform of the Weyl potential on the hypersurface at time~$\eta$.

In the linear regime, or more generally whenever the unequal-time Weyl-potential spectrum factorises, we may write $\Phi_W(\k,\eta_0-\chi') = g_\Phi(\eta_0-\chi')\,\Phi_W(\k)$, so that
\begin{equation}
    \psi^{\rm(O)}(\k; \chi, \hat{\n})
    = \mathcal{I}_\psi(\k; \chi, \hat{\n})\,\Phi_W(\k),
\end{equation}
where the line-of-sight kernel is
\begin{equation}
    \mathcal{I}_\psi(\k; \chi, \hat{\n})
    = \int_0^\chi \diff\chi'\, W_L(\chi, \chi')\,
    g_\Phi(\eta_0-\chi')\, e^{i k\mu \chi'}.
\end{equation}
This may be compared with the standard source-space Fourier transform of the lensing potential for a fixed observer at the origin which is considerably more complicated. Consider
\begin{equation}
    \tilde\psi(\k)
    =
    \int \diff^3 x_s \, e^{-i\k\cdot\x_s}\,
    \psi(\bm{0};\chi_s,\hat{\n}),
\end{equation}
with $\x_s=\chi_s\hat{\n}$ promoted to the integration variable. For a formal FT the integral is over $\mathbb{R}^3$, but here the point $\x_s$ is on the lightcone of the observer at $\x_0=0$ rather than a point on a spatial slice. 
Using
\begin{equation}
    \psi(\bm{0};\chi_s,\hat{\n})
    =
    \int_0^{\chi_s} \diff\chi'\, W_L(\chi_s,\chi')\,
    \Phi_W(\chi'\hat{\n},\eta_0-\chi')\,,
\end{equation}
and expanding $\Phi_W$ in its own Fourier modes then gives a convolution over internal momenta along the line of sight:
\begin{align}
    \tilde\psi(\k)
    &=
    \int \frac{\diff^3 k'}{(2\pi)^3}
    \int \diff\Omega \int_0^\infty \diff\chi_s\, \chi_s^2\,
    e^{-i\k\cdot\hat{\n}\chi_s} \nonumber\\ 
    &\int_0^{\chi_s} \diff\chi'\, W_L(\chi_s,\chi')\,
    e^{i\bm{k}'\cdot\hat{\n}\chi'}\,
    \Phi_W(\bm{k}',\eta_0-\chi').
\end{align}
Thus the source-space transform couples the external mode $\k$, conjugate to the source endpoint, to the internal modes $\k'$ of the Weyl potential along the line of sight. In the observer transform, by contrast, the same observer displacement is common to every point on the ray, so the line-of-sight integral acts as a transfer function on a single external mode.

Going back to the OFT, in the separable linear case, statistical homogeneity then gives, for the observer power spectrum,
\begin{equation}
    \Pobs_\psi(\k; \chi_1, \chi_2, \hat{\n}_1, \hat{\n}_2)
    = \mathcal{I}_\psi(\k; \chi_1, \hat{\n}_1)\,
      \mathcal{I}_\psi^*(\k; \chi_2, \hat{\n}_2)\,
      P_\Phi(k),
    \label{eq:Pobs_separable}
\end{equation}
where $P_\Phi(k)$ is the standard power spectrum of the reference Weyl potential. Crucially, $\Pobs_\psi$ is \emph{proportional} to $P_\Phi(k)$~-- the line-of-sight integration generates transfer functions $\mathcal{I}_\psi$ acting on the external legs, but does not introduce geometrical mode-mixing or convolutions over external Fourier modes. Physical observables such as the convergence $\kappa = -\tfrac{1}{2}\nabla_{\hat{\n}}^2 \psi$ then involve angular derivatives acting on the phase factor $e^{ik\mu\chi'}$ inside $\mathcal{I}_\psi$.\footnote{The lensing-potential kernel $W_L \sim 1/\chi'$ contains a logarithmic divergence as $\chi' \to 0$, reflecting unobservable angular monopole and dipole pieces associated with the observer endpoint. Physical observables like the convergence involve angular derivatives. Applying the unit-sphere Laplacian to the phase gives $\nabla_{\hat{\n}}^2 e^{i k \mu \chi'} = [-k^2 \chi'^2 (1-\mu^2) - 2ik\chi'\mu] e^{i k \mu \chi'}$. Multiplying by $W_L\sim1/\chi'$ gives terms proportional to $\chi'$ and to a finite constant as $\chi'\to0$, so the convergence kernel is regular at the observer.} As in the local case, wide-separation effects are fully taken into account.

\subsubsection{Non-linear observables}

The examples above were written at linear order, using the background lightcone  $(\chi,\hat{\n})$.  At second and higher orders, observables contain source, path and observer contributions, including nonlinear local terms, perturbations to the redshift-distance relation, lens-lens couplings, time-delay terms and various post-Born corrections. These terms change any observable inserted into the observer Fourier transform, but not the definition of the transform or power spectrum itself which is defined for fields on the background. Perturbed lightcone effects enter through the observable under consideration, not the formation of the statistic. 

To see the general structure at second order as an example, we can write the observer Fourier transform up to second order as
\begin{equation}
  X^{\rm(O)}(\k;\chi,\hat{\n}) =
  X^{\rm(O)}_{(1)}(\k;\chi,\hat{\n})
  + X^{\rm(O)}_{(2)}(\k;\chi,\hat{\n})+\cdots ,
\end{equation}
with
$%\begin{equation}
  X^{\rm(O)}_{(1)}(\k;\chi,\hat{\n})
  =
  F^{\rm(O)}_1(\k;\chi,\hat{\n})\,\delta(\k),
$%\end{equation}
and
\begin{equation}
  X^{\rm(O)}_{(2)}(\k;\chi,\hat{\n})
  =
  {1\over 2}
  \int { \diff^3 q\over (2\pi)^3}\,
  F^{\rm(O)}_2(\q,\k-\q;\chi,\hat{\n})\,
  \delta(\q)\delta(\k-\q).
  \label{eq:X2_observer_kernel}
\end{equation}
The kernel $F^{\rm(O)}_2$ is the full second-order observer-space kernel. For a local source contribution it reduces to the translated source kernel
\begin{equation}
  F^{\rm(O)}_2(\q,\k-\q;\chi,\hat{\n})
  =
  e^{ik\mu\chi}
  F_2(\q,\k-\q;\eta_0-\chi,\hat{\n})\,.
  \label{eq:F2_local_phase}
\end{equation}
(Here the left-hand side is labelled by the lightcone coordinate $\chi$, while the source kernel on the right is evaluated on the spatial hypersurface containing the source, at conformal time $\eta=\eta_0-\chi$.)
For integrated and post-Born terms the fields are evaluated at different points along the ray, so $F^{\rm(O)}_2$ contains line-of-sight integrals with phases depending separately on the internal momenta, such as $e^{i\q\cdot\hat{\n}\chi'}e^{i(\k-\q)\cdot\hat{\n}\chi''}$, rather than a single source phase $e^{ik\mu\chi}$.

The observer power spectrum has the usual expansion,
\begin{equation}
  P_X^{\rm(O)}
  =
  P^{\rm(O)}_{11}
  + P^{\rm(O)}_{12}
  + P^{\rm(O)}_{21}
  + P^{\rm(O)}_{22}
  + \cdots .
\end{equation}
For Gaussian linear fields the mixed terms $P^{\rm(O)}_{12}$ and $P^{\rm(O)}_{21}$ vanish, $P^{\rm(O)}_{11}$ is as above,  while for a local second-order source contribution the phase relation in Eq.~\eqref{eq:F2_local_phase} gives
\begin{align}
  &P^{\rm(O)}_{22,\rm loc}(\k;\chi_1,\chi_2,\hat{\n}_1,\hat{\n}_2)
  =
  e^{i\k\cdot(\chi_1\hat{\n}_1-\chi_2\hat{\n}_2)} \nonumber\\ \nonumber
  &\int { \diff^3 q\over 2(2\pi)^3}\,
  F_2(\q,\k-\q;\eta_1,\hat{\n}_1)
  F_2^*(\q,\k-\q;\eta_2,\hat{\n}_2) \\ 
  &P_\delta(q;\eta_1,\eta_2)P_\delta(|\k-\q|;\eta_1,\eta_2).
  \label{eq:P22_local_kernel}
\end{align}
For integrated and post-Born contributions there is no analogous single source phase: the two fields in the quadratic term may be evaluated at different positions on the ray, so their phases depend separately on $\q$ and $\k-\q$ inside the line-of-sight integrals. These terms are therefore naturally included in the full observer-space kernel $F^{\rm(O)}_2$, rather than in a source kernel multiplied by $e^{ik\mu\chi}$. The distinction affects the integrand of $P^{\rm(O)}_{22}$, but not the overall momentum-conserving delta function. That delta function follows from translating the observer position on the hypersurface, so it is present independently of whether the observable is linear, local, integrated, or evaluated beyond the Born approximation.

\subsubsection{Double-Legendre expansion}
\label{sec:double_legendre}

For a fixed scalar observable $X$, rotational invariance implies that the dependence of its observer spectrum on the direction of $\k$ can be written in terms of $\mu_1=\hat{\k}\cdot\hat{\n}_1$, $\mu_2=\hat{\k}\cdot\hat{\n}_2$ and $\nu=\hat{\n}_1\cdot\hat{\n}_2$. The last of these is fixed by the choice of lightcone directions, which are held fixed in the observer spectrum. We can expand this in a double-Legendre series,
\be\label{eq:double_legendre_def}
  \Pobs(k;\chi_1,\chi_2,\mu_1,\mu_2,\nu)
  = \sum_{\ell_1,\ell_2}
    P_{\ell_1\ell_2}(k;\chi_1,\chi_2,\nu)\,
    \mathcal{L}_{\ell_1}(\mu_1)\,\mathcal{L}_{\ell_2}(\mu_2)\,.
\ee
Since $X$ is real-valued, the OFT obeys
$X^{(\mathrm{O})}(-\k)=X^{(\mathrm{O})*}(\k)$, so that
$\Pobs(k,-\mu_1,-\mu_2)=[\Pobs(k,\mu_1,\mu_2)]^*$. This implies
\be\label{eq:parity_Pll}
  P_{\ell_1\ell_2}^*
  = (-1)^{\ell_1+\ell_2}\,P_{\ell_1\ell_2},
\ee
which implies the coefficients are real when $\ell_1+\ell_2$ is even and purely
imaginary when $\ell_1+\ell_2$ is odd. For local fields
(Eq.~\ref{eq:Pobs_local_result}), the imaginary part originates from
the sine of the geometric phase
$e^{ik(\mu_1\chi_1-\mu_2\chi_2)}$, which is non-zero whenever the
two legs are unequal, as well as from $P$ from contributions such as the Doppler term. For
integrated observables (Eq.~\ref{eq:Pobs_separable}), the same
mechanism operates inside the line-of-sight integrals.

Integrating over $\hat{\k}$ and applying the addition theorem defines
the Fourier angle-averaged observer spectrum,
\be\label{eq:Pobs_angle_averaged}
  \barPobs(k;\chi_1,\chi_2,\nu)
  \equiv \int\frac{\diff\Omega_k}{4\pi}\,\Pobs
  = \sum_\ell \frac{P_{\ell\ell}(k;\chi_1,\chi_2,\nu)}{2\ell+1}\,
    \mathcal{L}_\ell(\nu)\,.
\ee
Only the diagonal multipoles $P_{\ell\ell}$ remain after this average; by
Eq.~\eqref{eq:parity_Pll} these are always real. For the separable
kernels that arise at linear order,
$P_{\ell_1\ell_2}=\K_{\ell_1}\K_{\ell_2}^*\,P_0(k)$ is an outer
product, so the off-diagonal multipoles carry no independent
information beyond what is contained in the set $\{\K_\ell\}$, and
there is no $\nu$-dependence in $P_{\ell\ell}$, but this is not the case more generally. 

\subsubsection{Plane-parallel multipoles}

The general form for $P^{(\mathrm{O})}$ contains all wide-angle and integrated effects without approximation, but it is useful to show how the usual plane-parallel Legendre expansion is recovered. We illustrate how to calculate the wide-angle expansion from $\Pobs$ in the Appendix.

In the plane-parallel limit both lines of sight coincide, $\hat{\bm{n}}_1 = \hat{\bm{n}}_2 \equiv \hat{\bm{n}}$. Typically both sources lie at the same comoving distance, $\chi_1 = \chi_2 \equiv \bar\chi$, but they do not have to. In this case the observer spectrum becomes
\be
  P^{(\mathrm{O})}_{\mathrm{PP}}(k, \mu; \chi_1, \chi_2)
  = \sum_{\ell_1, \ell_2} P_{\ell_1\ell_2}(k; \chi_1, \chi_2,\nu=1)\,
    \mathcal{L}_{\ell_1}(\mu)\, \mathcal{L}_{\ell_2}(\mu)\,.
\ee
The standard redshift-space power spectrum multipoles, defined by
\be\label{redshftmultipoles}
  P_L(k, \chi_1, \chi_2) = \frac{2L+1}{2}\int_{-1}^{1} \diff\mu\, \mathcal{L}_L(\mu)\, P^{(\mathrm{O})}_{\mathrm{PP}}(k, \mu; \chi_1, \chi_2)\,,
\ee
follow as 
\be\label{eq:PL_3j}
  P_L(k;  \chi_1, \chi_2) = \sum_{\ell_1, \ell_2} (2L+1)
  \begin{pmatrix} \ell_1 & \ell_2 & L \\ 0 & 0 & 0 \end{pmatrix}^{\!2}
  P_{\ell_1\ell_2}(k; \chi_1, \chi_2,\nu=1)\,.
\ee
The $3j$ symbol enforces $|\ell_1{-}\ell_2| \le L \le \ell_1{+}\ell_2$ and $\ell_1{+}\ell_2{+}L$ even.

For a local observable, the phases cancel when $\chi_1 = \chi_2$ and $\hat{\bm{n}}_1 = \hat{\bm{n}}_2$ (Sec.~\ref{sec:local_field_limit}), reducing $P^{(\mathrm{O})}_{\mathrm{PP}}$ to the standard equal-time power spectrum $P(k,\mu;\bar\eta)$ on the hypersurface at $\bar\eta = \eta_0 - \bar\chi$. If $P$ is a finite polynomial in $\mu$~-- as for density plus Kaiser RSD, where $P_\Delta \propto (b+f\mu^2)^2$ is quartic~-- then only finitely many $P_{\ell_1\ell_2}$ contribute and the $3j$ sum truncates to $L = 0,2,4$. This is not the case when $\chi_1 \neq \chi_2$, and odd multipoles are non-zero~-- see below, Fig.~\ref{fig:P_L_plot}.

For integrated observables such as the lensing potential
(Sec.~\ref{sec:lensing_integrated}), the phase $e^{ik\mu\chi'}$ is
evaluated at the integration variable $\chi'$, not at $\bar\chi$. In
the product $|\mathcal{I}|^2$, as in the equal-time case, the phase becomes
$e^{ik\mu(\chi'-\chi'')}$, which does \emph{not} cancel since the two
integration variables range independently over $[0,\bar\chi]$.
Consequently $P^{(\mathrm{O})}_{\mathrm{PP}}$ is not a polynomial in
$\mu$ and the $3j$ sum does not truncate: integrated observables and
their cross-terms with local effects generate contributions at all
multipoles $L$, with the parity constraint of
Eq.~\eqref{eq:parity_Pll} determining which are real and which
imaginary. For an equal-distance auto-spectrum of a real observable,
the final result is real and even in $\mu$, so only even $L$ survive.
For unequal radial distances, or for asymmetric cross-correlations, the
odd multipoles need not vanish and are fixed by the Hermiticity condition
in Eq.~\eqref{eq:parity_Pll}.

\subsection{Two-Point Correlation Functions}

The two-point correlation function is usually defined for a single observer, correlating measurements on that observer's past lightcone (the `lightcone 2pcf'). In the OFT framework, however, there is a second, equally natural object: the correlation between measurements made by two different observers. This is the configuration-space counterpart of the observer power spectrum.

\subsubsection{The Two-Observer Correlation Function}

Because the OFT is defined by transforming over observer positions $\bm{x}_0$ at fixed lightcone coordinates, the fundamental configuration-space statistic is the correlation between two observers. We define the \emph{observer-space correlation function} as
\begin{equation}
  \Xi(\bm{y}; \chi_1, \hat{\bm{n}}_1; \chi_2, \hat{\bm{n}}_2)
  \equiv
  \big\langle
    X(\bm{x}_0; \chi_1, \hat{\bm{n}}_1)\,
    X(\bm{x}_0+\bm{y}; \chi_2, \hat{\bm{n}}_2)
  \big\rangle,
\end{equation}
where $\bm{y} = \bm{x}_0' - \bm{x}_0$ is the separation between two observers: the first measures the lightcone leg $(\chi_1,\hat{\bm n}_1)$ and the second, displaced by $\bm y$, measures $(\chi_2,\hat{\bm n}_2)$. Statistical homogeneity implies that $\Xi$ is independent of the absolute observer position and depends on the observers only through their separation vector $\bm y$; isotropy then restricts this dependence to $|\bm y|$ and its orientation relative to the fixed lines of sight, i.e. through $\hat{\bm y}\cdot\hat{\bm n}_i$ and $\hat{\bm n}_1\cdot\hat{\bm n}_2$.
Simplifying the two-observer correlation gives, using the OFT,
\begin{equation}\label{eq:Xi_from_Pobs}
  \Xi(\bm{y}; \chi_1,\hat{\bm{n}}_1; \chi_2,\hat{\bm{n}}_2)
  = \int \frac{\diff^3 k}{(2\pi)^3}\,
      e^{-i\bm{k}\cdot\bm{y}}\,
      P^{\rm(O)}(\bm{k}; \chi_1,\chi_2,\hat{\bm{n}}_1,\hat{\bm{n}}_2).
\end{equation}
Therefore, the observer power spectrum is simply the Fourier transform of this two-observer correlation with respect to the observer separation:
\begin{equation}
  P^{\rm(O)}(\bm{k}; \chi_1, \chi_2, \hat{\bm{n}}_1, \hat{\bm{n}}_2)
  =
  \int \diff^3 y \, e^{i\bm{k}\cdot\bm{y}}\,
    \Xi(\bm{y}; \chi_1, \hat{\bm{n}}_1; \chi_2, \hat{\bm{n}}_2).
\end{equation}
This is simply analogous to the usual relation between the equal-time correlation function $\xi(\bm{s})$ and the power spectrum $P(\bm{k})$, but with the roles of field point and observer position interchanged. Here $P^{\rm(O)}$ encodes correlations between light-rays on different lightcones and in different directions, rather than between points on a single spatial slice.

\subsubsection{The Single-Observer Lightcone Correlation Function}

The standard observable in galaxy surveys is the correlation function measured by a single observer. Fixing the observer at the origin, $\bm{x}_0 = \bm{0}$, the lightcone field is $X(\bm{0}; \chi, \hat{\bm{n}})$, which we express using the inverse OFT:
\begin{equation}\label{ckdsjnsdkjcnskjc}
  X(\bm{0}; \chi, \hat{\bm{n}})
  =
  \int \frac{\diff^3 k}{(2\pi)^3}\,
    X^{\rm(O)}(\bm{k}; \chi, \hat{\bm{n}}).
\end{equation}
The single-observer lightcone two-point function is therefore 
\begin{align}
  \xi_{\rm LC}(\chi_1, \hat{\bm{n}}_1; &\chi_2, \hat{\bm{n}}_2)
  \equiv
  \big\langle
    X(\bm{0}; \chi_1, \hat{\bm{n}}_1)\,
    X(\bm{0}; \chi_2, \hat{\bm{n}}_2)
  \big\rangle
  \nonumber \\
  &=
  \int \frac{\diff^3 k_1}{(2\pi)^3}
       \frac{\diff^3 k_2}{(2\pi)^3}\,
  \big\langle
    X^{\rm(O)}(\bm{k}_1; \chi_1, \hat{\bm{n}}_1)\,
    X^{\rm(O)}(\bm{k}_2; \chi_2, \hat{\bm{n}}_2)
  \big\rangle
  \nonumber \\
  &=
  \int \frac{\diff^3 k}{(2\pi)^3}\,
    P^{\rm(O)}(\bm{k}; \chi_1, \chi_2, \hat{\bm{n}}_1, \hat{\bm{n}}_2),
  \label{eq:xi_from_Pobs}
\end{align}
Thus the single-observer lightcone correlation function is just the momentum-space integral of the observer power spectrum. Equivalently,
\begin{equation}
  \xi_{\rm LC}(\chi_1,\hat{\bm{n}}_1; \chi_2,\hat{\bm{n}}_2)
  =
  \Xi(\bm{y}=\bm{0}; \chi_1,\hat{\bm{n}}_1; \chi_2,\hat{\bm{n}}_2),
\end{equation}
the zero-separation limit of the two-observer correlation. While $\Xi(\bm{y})$ describes how measurements correlate between different observers separated by $\bm{y}$, $\xi_{\rm LC}$ describes correlations for the same observer viewing different points on their lightcone.

For local observables (density, Kaiser RSD, etc.) the observer spectrum
factorises into a geometric phase $e^{i\bm{k}\cdot\bm{s}}$, where $\bm{s}=\bm{x}_1-\bm{x}_2$, times the standard unequal-time $P(\bm{k};\eta_1,\eta_2,\hat{\bm{n}}_1,\hat{\bm{n}}_2)$, giving
\begin{equation}
  \xi_{\rm LC}(\chi_1,\hat{\bm{n}}_1;\chi_2,\hat{\bm{n}}_2)
  =
  \int \frac{\diff^3 k}{(2\pi)^3}\,
    e^{i\bm{k}\cdot\bm{s}}\,
    P(\bm{k};\eta_1,\eta_2,\hat{\bm{n}}_1,\hat{\bm{n}}_2),
  \label{eq:xi_local_from_PX}
\end{equation}
which is just the standard relation between the unequal-time 2-point
correlation function of a local field and its (possibly anisotropic) power
spectrum. In the equal-time, plane-parallel limit ($\eta_1=\eta_2$,
$\hat{\bm{n}}_1\simeq\hat{\bm{n}}_2\equiv\hat{\bm{n}}$),
Eq.~\eqref{eq:xi_local_from_PX} reduces to the usual
$\xi(\bm{s})=\xi(s,\mu)$ with $\mu=\hat{\bm{k}}\cdot\hat{\bm{n}}$ and
$P(\bm{k})=P(k,\mu)$ the constant-time redshift-space power spectrum.

\subsubsection{Which 2-point correlation function?}

The usual lightcone 2pcf is the most closely related to observations; it is obtained by integrating the observer power spectrum over its Fourier mode. The two-observer viewpoint, on the other hand, is more natural for cosmological simulations. In a periodic $N$-body box one can place the apex of the lightcone at many different grid points, generating an ensemble of observers and associated light-rays. The two-observer correlation $\Xi(\bm{y})$ then quantifies how lightcone measurements at fixed relative coordinates $(\chi_i,\hat{\bm{n}}_i)$ correlate as a function of the separation $\bm{y}$ between observer locations. Its Fourier transform gives $P^{\rm(O)}$ directly, without any geometric projection or survey window. This makes the observer spectrum a natural theoretical object for validation against simulations: it can be measured from the full ensemble of observers (for each pair $(\chi_i,\hat{\bm{n}}_i)$) and only afterwards reduced to the standard lightcone correlation or its multipole decompositions~\citep{2021MNRAS.501.2547G}.

\subsection{Lightcone Power Spectra}

While the observer power spectrum preserves translation invariance, several alternative Fourier-space formalisms are normally used which break translation invariance and so give non-diagonal or mixed-space representations of the fields. We now show that these conventional lightcone power spectra are just integrals over $P^{\rm(O)}$. (Throughout we ignore any survey window, working with idealised all-space (full-sky) fields for simplicity.) 

\subsubsection{The lightcone power spectrum}

The standard approach is to Fourier transform the lightcone field itself, defined with respect to a fixed observer at the origin. For purely local observables this distinction is often hidden by working on an approximate spatial slice, but for evolving or integrated lightcone observables the choice of integration domain is essential. We define the lightcone Fourier transform (LCFT) as
\begin{equation}
  X_{\LC}(\bm{k}) \equiv \int_{\LC} \diff^3 x_s \, e^{-i\bm{k}\cdot\bm{x}_s} \,
  X(\bm{0}; \chi_s, \hat{\bm{n}}),
\end{equation}
where the integration is over the observer's past lightcone and
$\diff^3 x_s = \chi_s^2 \diff\chi_s \diff\Omega$ (we assume for simplicity an infinite lightcone). The two-point statistics
of $X_{\LC}$ define the lightcone power spectrum $\Plc$~\citep{1999ApJ...517....1Y,1999ApJ...527..488Y},
\begin{equation}
  \big\< X_{\LC}(\bm{k}_1) X_{\LC}^*(\bm{k}_2) \big\> \equiv
  \Plc(\bm{k}_1, \bm{k}_2),
\end{equation}
which is in general non-diagonal in Fourier space due to mode mixing (this is the covariance of the field). It becomes diagonal only when the lightcone reduces to a single constant-time slice, so that $X(\bm{0};\chi_s,\hat{\bm n})$ is a statistically homogeneous field and $X_{\rm LC}(\bm{k})$ is just the usual Fourier transform.

To relate this to the observer spectrum, we express the lightcone field
$X(\bm{0}; \chi, \hat{\bm{n}})$ using the inverse observer Fourier
transform, Eq. \eqref{ckdsjnsdkjcnskjc}.
%\begin{equation}
%  X(\bm{0}; \chi, \hat{\bm{n}})
%  =
%  \int \frac{\diff^3 k}{(2\pi)^3} \, X^{\rm(O)}(\bm{k}; \chi, \hat{\bm{n}}).
%\end{equation}
Substituting this into the definition of the LCFT and taking the ensemble
average, we obtain the projection formula
\begin{align}
  \Plc(\bm{k}_1, \bm{k}_2)
  &=
  \int \frac{\diff^3 k}{(2\pi)^3}
    \int_{\LC} \diff^3 x_{s1}
    \int_{\LC} \diff^3 x_{s2} \,
    e^{-i\bm{k}_1\cdot\bm{x}_{s1}} \,
    e^{+i\bm{k}_2\cdot\bm{x}_{s2}} \, \\ \nonumber
    &P^{\rm(O)}\!\big(\bm{k}; \chi_{s1}, \chi_{s2},
    \hat{\bm{n}}_1, \hat{\bm{n}}_2\big).
  \label{eq:Plc_projection}
\end{align}
This makes it explicit that the non-diagonality of
$\Plc(\bm{k}_1, \bm{k}_2)$ is purely geometric. The underlying
statistics $P^{\rm(O)}(\bm{k})$ are diagonal in the observer wavenumber
$\bm{k}$; mode mixing in $\Plc$ is generated only by projecting these
modes through the conical geometry of a single observer's past lightcone.

\subsubsection{The mixed space power spectrum}

In wide-angle RSD analyses, it is common to
define a mixed-space spectrum $\Pkd(\bm{k}, \bm{d})$ that depends on both
a Fourier mode $\bm{k}$ and a configuration-space pair centre $\bm{d}$~\citep{1998ApJ...498L...1S,2000ApJ...535....1M,2004ApJ...614...51S,2008MNRAS.389..292P,2016JCAP...01..048R,2018MNRAS.476.4403C,2019JCAP...03..040B,2022JCAP...01..061C,2023JCAP...04..067P,2026JCAP...06..039A}. The Fourier component is then a transform over the separation vector between the points,
 $\bm{s} = \bm{x}_1 - \bm{x}_2$ on
the lightcone and $\bm{d} = (\bm{x}_1+\bm{x}_2)/2$ is their midpoint. Other definitions of $\bm d$, such as the endpoint $\bm{d} = \bm{x}_1$, are often used, but for simplicity we do not consider them here. More concretely, the
mixed spectrum is defined as the Fourier transform of the lightcone
correlation function with respect to the separation $\bm{s}$ while holding
the pair centre $\bm{d}$ fixed:
\begin{equation}
  \Pkd(\bm{k}, \bm{d}) \equiv
  \int \diff^3 s \, e^{-i\bm{k}\cdot\bm{s}} \,
  \xi_{\LC}\!\left( \bm{d}+\frac{\bm{s}}{2},
                    \bm{d}-\frac{\bm{s}}{2} \right).
\end{equation}
Substituting the expression for $\xi_{\LC}$ in terms of
$P^{\rm(O)}$ from Eq.~\eqref{eq:xi_from_Pobs}, we find the mixed-space spectrum as an integral transform over the observer power spectrum,
\begin{equation}
  \Pkd(\bm{k}, \bm{d}) =
  \int \frac{\diff^3 k'}{(2\pi)^3}
  \int \diff^3 s \, e^{-i\bm{k}\cdot\bm{s}} \,
  P^{\rm(O)}\!\big(\bm{k}';
  \chi_1, 
  \chi_2,
  \hat{\bm{n}}_1, \hat{\bm{n}}_2)
  \label{eq:Pkd_exact}
\end{equation}
where the lightcone coordinates
$\{\chi_i, \hat{\bm{n}}_i\}$ are implicit functions of the centre
$\bm{d}$ and separation $\bm{s}$. This relation holds for any observable field $X$. This can be simplified to
\begin{equation}
  \Pkd(\bm{k}, \bm{d}) =
  \sum_{\ell}\frac{1}{{2\ell+1}}
  \int \frac{k'^2 \diff k'}{2\pi^2}
  \int \diff^3 s \, e^{-i\bm{k}\cdot\bm{s}} \,
  {P_{\ell\ell}(k';\chi_1,\chi_2,\nu)}
  \mathcal{L}_\ell(\nu),
  \label{eq:Pkd_from_Pll}
\end{equation}
in terms of the Legendre multipoles of $P^{\rm(O)}$.

The mixed spectrum and the non-diagonal lightcone kernel are just alternative
projections of the observer spectrum. Defining
$\bm{x}_{1,2}=\bm d\pm \bm s/2$, with
$\bm{k} \equiv (\bm{k}_1+\bm{k}_2)/2$ and
$\Delta\bm{k} \equiv \bm{k}_1-\bm{k}_2$,
$\tilde P$ and $\Pkd$ are Fourier transforms of one another with respect
to the pair centre:
\begin{align}
  &\tilde P\!\left(\bm{k}+\frac{\Delta\bm{k}}{2},\,\bm{k}-\frac{\Delta\bm{k}}{2}\right)
  =
  \int \diff^3 d \, e^{-i\Delta\bm{k}\cdot\bm d}\,\Pkd(\bm{k},\bm d),
  \qquad \\
  &\Pkd(\bm{k},\bm d)
  =
  \int \frac{\diff^3 \Delta k}{(2\pi)^3}\,
  e^{i\Delta\bm{k}\cdot\bm d}\,
  \tilde P\!\left(\bm{k}+\frac{\Delta\bm{k}}{2},\,\bm{k}-\frac{\Delta\bm{k}}{2}\right).
  \label{eq:Pkd_tildeP_relation}
\end{align}
For local observables in the distant-observer, slowly-varying limit, the
$\bm s$-integration approximately selects $\bm{k}'\simeq\bm{k}$, so
$\Pkd(\bm{k},\bm d)$ behaves like a quasi-local power spectrum at pair
centre $\bm d$, which is why a wide-angle expansion is used in standard Kaiser RSD
analyses. More generally, however, both $\Pkd$ and $\tilde P$ are
complicated lightcone projections of $P^{\rm(O)}$~-- which means that wide-angle geometry
makes the dependence on $\bm d$ and $\bm s$ non-trivial, wide separations
break the slowly-varying approximation, and integrated observables such
as lensing are intrinsically non-local along the line of sight and combine into a complicated convolution.

\subsubsection{Connection to the Yamamoto estimator}
\label{sec:yamamoto}

The Yamamoto estimator measures redshift-space power spectrum multipoles
from galaxy surveys~\citep{2006PASJ...58...93Y,2015MNRAS.453L..11B,2015PhRvD..92h3532S}. Using the FKP field
$F(\bm x)=w(\bm x)\,[n_g(\bm x)-\alpha n_r(\bm x)]$, the $\ell$th
multipole estimator is
\begin{equation}
  \hat P_\ell(k)
  =
  \frac{2\ell+1}{I}
  \int \frac{\diff\Omega_k}{4\pi}
  \int \diff^3x_1 \diff^3x_2\,
  e^{-i\bm k\cdot\bm s}\,
  \mathcal L_\ell(\hat{\bm k}\cdot\hat{\bm x}_1)\,
  F(\bm x_1)F(\bm x_2),
\end{equation}
where $\bm s=\bm x_1-\bm x_2$ and $I=\int \diff^3x\,\bar n^2 w^2$.
The Legendre polynomial is evaluated at the endpoint
$\hat{\bm x}_1$ rather than the pair midpoint
$\hat{\bm d}=\widehat{(\bm x_1+\bm x_2)/2}$; the two conventions
differ at order $s/d$ and agree in the plane-parallel limit. The
endpoint choice is standard in practice because it makes the estimator
factorisable into FFTs.

Neglecting shot noise, and substituting
$\xi_{\rm LC}=\int \diff^3k'/(2\pi)^3\,P^{\rm(O)}$ from
Eq.~\eqref{eq:xi_from_Pobs}, the expectation value becomes
\begin{align}\label{eq:yamamoto_from_Pobs}
  \big<\hat P_\ell(k)\big>
  &=
  \frac{2\ell+1}{I}
  \int \frac{\diff\Omega_k}{4\pi}
  \int \frac{\diff^3k'}{(2\pi)^3}
  \int \diff^3x_1 \diff^3x_2\,
  e^{-i\bm k\cdot\bm s}\,
  \mathcal L_\ell(\hat{\bm k}\cdot\hat{\bm x}_1)\, \\ \nonumber
  &W(\bm x_1)W(\bm x_2)\,
  P^{\rm(O)}(\bm k';\chi_1,\chi_2,\hat{\bm n}_1,\hat{\bm n}_2),
\end{align}
with $W=\bar n w$. The estimator is therefore a windowed projection of
the observer spectrum. For local observables, the observer spectrum
contributes a phase $e^{i\bm k'\cdot\bm s}$ which, in the
distant-observer slowly varying window limit, selects $\bm k'\approx\bm k$ and recovers the
standard window-convolved multipoles. For integrated observables such
as lensing, the $\bm k'$ dependence sits inside the line-of-sight
kernels and this simplification does not occur.
In Eq.~(\ref{eq:yamamoto_from_Pobs}), it remains exact in both cases, providing a complete model for the estimator that includes wide-angle, evolution and integrated effects~\citep{2022JCAP...01..061C,2020JCAP...07..048B,2023PhRvL.131k1201F}.

\subsubsection{Comparison of the four power spectra}
\label{sec:four_spectra}

\begin{figure*}
\resizebox{0.9\textwidth}{!}{%
\centering
\begin{tikzpicture}[
    every node/.style={align=center},
    hub/.style={draw, rounded corners=4pt, fill=black!5,
                minimum width=3.2cm, minimum height=0.85cm,
                font=\normalsize, inner sep=4pt, thick},
    mid/.style={draw, rounded corners=3pt, fill=yellow!5,
                minimum width=2.6cm, minimum height=0.75cm,
                font=\normalsize, inner sep=3pt, thick},
    mid2/.style={draw, rounded corners=3pt, fill=red!5,
                minimum width=2.6cm, minimum height=0.75cm,
                font=\normalsize, inner sep=3pt, thick},
    outer/.style={draw, rounded corners=3pt, fill=green!5,
                  minimum width=2.6cm, minimum height=0.75cm,
                  font=\normalsize, inner sep=3pt, thick},
    approx1/.style={draw, rounded corners=3pt, fill=blue!5,
                  minimum width=2.6cm, minimum height=0.75cm,
                  font=\normalsize, inner sep=3pt, thick},
    arr/.style={-{Stealth[length=10pt]}, thick, black!80},
    grey/.style={<->, gray!50, thick, dashed},
    greyarr/.style={-{Stealth[length=10pt]}, black!60, thick, dashed},
    lbl/.style={font=\scriptsize, fill=white, inner sep=2pt, sloped},
    glbl/.style={font=\scriptsize, fill=white, inner sep=1.5pt,
                 text=black!60}
  ]

  % ---- Central hub ----
  \node[hub] (PO) at (0,0)
    {$P^{\rm(O)}(\k;\chi_1,\chi_2,\hat{\n}_1,\hat{\n}_2)$};
  % ---- Central hub ----
  \node[approx1] (P) at (0,2.8)
    {$P(\k;\chi_1,\chi_2,\hat{\n}_1,\hat{\n}_2)$};
  \node[approx1] (PL) at (0,5)
    {$P_L(k, \chi_1, \chi_2)$};
  \node[approx1] (PP) at (0,7)
    {$P(k,\mu, \chi)$};

  % ---- Left branch: angular decomposition ----
  \node[mid2] (Pl1l2) at (-7, 2.8)
    {$P^{\rm(O)}_{\ell_1\ell_2}(k;\chi_1,\chi_2)$};
  \node[mid2] (Pll) at (-7, 0)
    {$P^{\rm(O)}_{\ell\ell}(k;\chi_1,\chi_2)$};
  \node[mid2] (Cell) at (-7, -2.8)
    {$C_\ell(\chi_1,\chi_2)$};
  \node[mid2] (SFB) at (-7, -5.6)
    {$C_\ell^\text{SFB}(k_1,k_2)$};

  % ---- Right branch: configuration space ----
  \node[mid] (XiY) at (7, 2.8)
    {$\Xi(\bm{y};\chi_1,\hat{\n}_1;\chi_2,\hat{\n}_2)$};
  \node[mid] (xiLC) at (7, 0)
    {$\xi_{\rm LC}(\chi_1,\hat{\n}_1;\chi_2,\hat{\n}_2)$};

  % ---- Bottom: survey pipeline ----
  \node[outer] (Pkd) at (4.5, -2.8)
    {$\bar{P}(\k,\bm{d})$};
  \node[outer] (Ptilde) at (1, -5.6)
    {$\tilde{P}(\k_1,\k_2)$};
  \node[outer] (Yam) at (7, -5.6)
    {$\langle\hat{P}_\ell(k)\rangle$};

  % ==== Main arrows from hub ====

  \draw[arr] (PO) -- (Pl1l2)
    node[lbl, pos=0.5] {Legendre decomposition in $\mu_i$};

  \draw[arr] (PO) -- (P)
    node[lbl, pos=0.5, right, sloped=false] {local field limit};
  \draw[arr] (P) -- (PL)
    node[lbl, pos=0.5, right, sloped=false] {plane parallel multipoles, $\hat\n_1=\hat\n_2$};
  \draw[arr] (PL) -- (PP)
    node[lbl, pos=0.5, right, sloped=false] {standard PP equal redshift $\chi_1=\chi_2$};

  \draw[arr] (PO) -- (XiY)
    node[lbl, pos=0.5] {inverse FT};

  \draw[arr] (PO) -- (xiLC)
    node[lbl, pos=0.5] {integrate over $\k$};

  % ==== Left branch downward ====
  \draw[arr] (Pl1l2) -- (Pll)
    node[lbl, pos=0.5, right, sloped=false] {angular average\\ over $\hat{\k}$};
  \draw[arr] (Pll) -- (Cell)
    node[lbl, pos=0.5, right, sloped=false] {integrate over $k$};
  \draw[arr] (Cell) -- (SFB)
    node[lbl, pos=0.5, right, sloped=false] {spherical Bessel transform in $\chi_i$};

  % ==== Right branch downward ====
  \draw[arr] (xiLC) -- (Pkd)
    node[lbl, pos=0.5] {FT over $\bm{s}$};
  \draw[arr] (Pkd) -- (Ptilde)
    node[lbl, pos=0.5] {FT over $\bm{d}$};
  \draw[arr] (Pkd) -- (Yam)
    node[lbl, pos=0.5] {survey };

  % ==== Grey (secondary) arrows ====

  \draw[greyarr] (XiY) -- (xiLC)
    node[glbl, pos=0.5, right] {$\bm{y}=\bm{0}$};

% Cell <-> xiLC
  \draw[greyarr]  (xiLC) -- (Cell)
    node[glbl, pos=0.7, sloped, left, above] {Legendre decomposition in $\nu$};

  \draw[greyarr]  (Pl1l2) -- (PL)
    node[glbl, pos=0.5, sloped, left, above] {Wigner sum, $\nu=1$};

  \draw[greyarr]  (Ptilde) -- (SFB)
    node[glbl, pos=0.5, sloped, above] {spherical Fourier Bessel};

  % PO to Ptilde direct
  \draw[greyarr] (PO.south) -- %++(0,-1.4) 
  (Ptilde.north)
    node[glbl, pos=0.55, sloped, above]
    {lightcone FT over $\bm{x}_i$};

\end{tikzpicture}
}
\caption{Hierarchy of lightcone two-point statistics. The observer
spectrum $P^{\rm(O)}$ (centre) is diagonal in $\k$ and is
the  quantity from which all other spectra are derived. (Blue, upwards; approximations) Successive approximations  give
the local-field spectrum $P$, its plane-parallel multipoles
$P_L(k;\chi_1,\chi_2)$, and finally the standard $P(k,\mu)$ in the
equal-redshift limit.  (Red, left; angular decompositions) Legendre decomposition in $\mu_1,\mu_2$ gives the
multipoles $P^{\rm(O)}_{\ell_1\ell_2}$, whose diagonal elements give
$C_\ell$ in configuration space after appropriate $\k$ integration which can be further expanded in spherical Bessel functions. 
(Right, yellow; configuration space) The inverse Fourier transform gives the two-observer
correlation $\Xi(\bm{y})$; and integration over $\k$ gives the
single-observer lightcone correlation $\xi_{\rm LC}$, which is also the
$\bm{y}{=}\bm{0}$ limit of $\Xi$.
(Lower right, green; lightcone power spectra)
The survey pipeline connects
$\xi_{\rm LC}$ through the mixed spectrum $\bar{P}(\k,\bm{d})$ to the
Yamamoto estimator $\langle\hat{P}_\ell\rangle$. The non-diagonal
lightcone kernel $\tilde{P}(\k_1,\k_2)$ is via another lightcone FT of
$\bar{P}$. Dashed arrows give some standard alternative transformations. }
\label{fig:hierarchy}
\end{figure*}
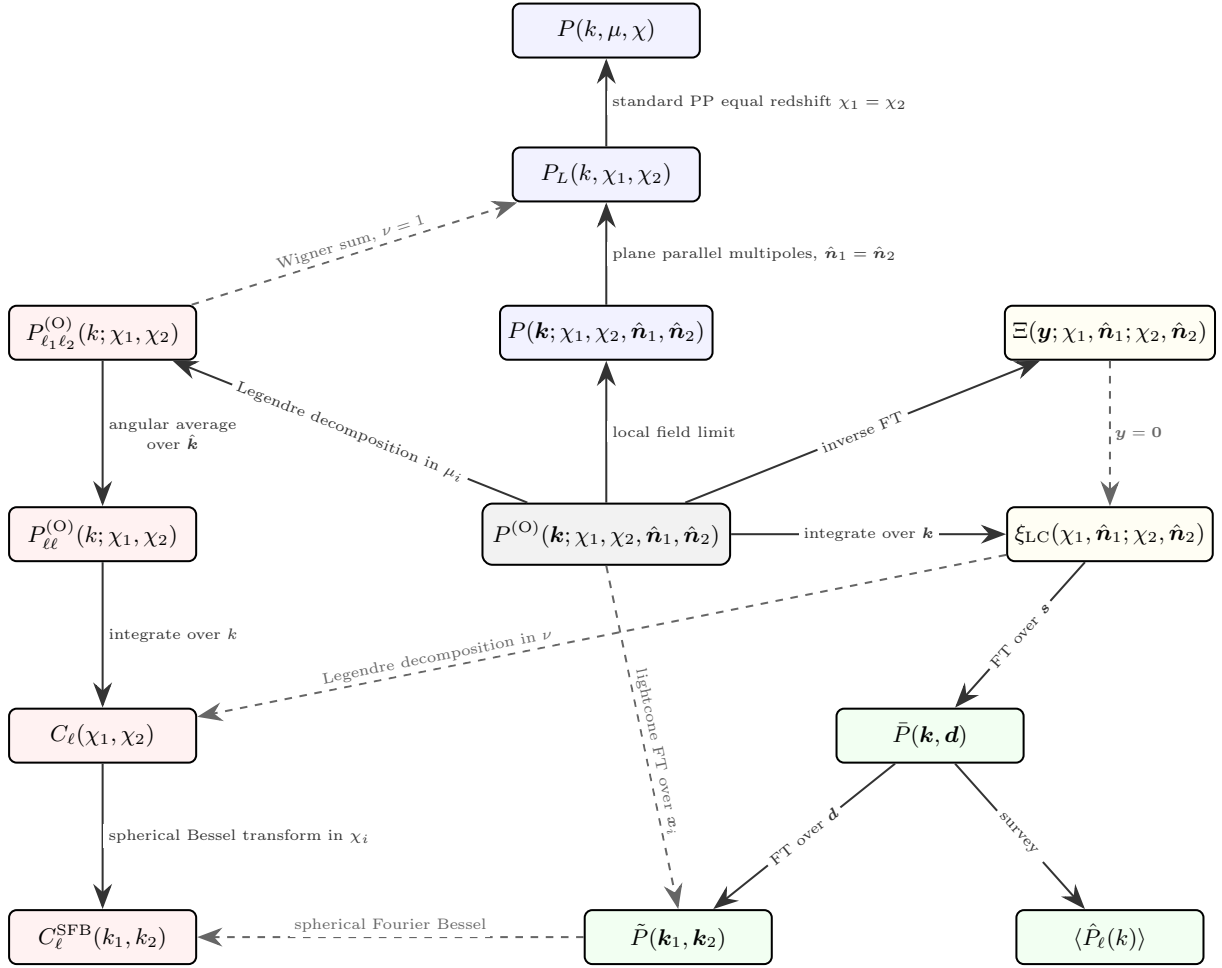

Four Fourier-space two-point statistics have now been defined which fall into two pairs.
The first pair~-- $P$ and~$\Pobs$ -- are diagonal in $\k$
and carry the lightcone coordinates $(\chi_i,\hat{\n}_i)$ as
free external parameters. The unequal-time spectrum
$P(\k;\eta_1,\eta_2,\hat{\n}_1,\hat{\n}_2)$ is defined on
spatial hypersurfaces and contains the full wide-angle
information \emph{for local observables}: the lines of sight
$\hat{\n}_1$, $\hat{\n}_2$ may subtend any opening angle, and
the lightcone correlation function follows from integrating
$e^{i\k\cdot\bm{s}}P$ over $\k$
(Eq.~\ref{eq:xi_local_from_PX}). However, $P$ is only
defined for fields that live on a spatial hypersurface; for
integrated observables such as lensing convergence it does not
exist. The observer spectrum $\Pobs$ extends the diagonal
description to any observable on the lightcone. For local fields
it is $e^{i\k\cdot\bm{s}}P$
(Eq.~\ref{eq:Pobs_local_result}), which has the same physical
content as $P$ but phase-shifted so that $C_\ell$, $\bar{P}$,
and the Yamamoto estimator are each obtained from it by
integration (Fig.~\ref{fig:hierarchy}). For integrated fields it
provides the only diagonal power spectrum function available.

The second pair -- $\Pkd(\k,\d)$ and $\Plc(\k_1,\k_2)$ --
are defined on a single observer's past lightcone rather than on
a spatial hypersurface. The mixed spectrum $\Pkd$ is the Fourier
transform of the lightcone correlation function over the pair
separation $\bm{s}$ at fixed midpoint $\d$; $\Plc$ results from
Fourier transforming the lightcone field over all source
positions. In both cases, the external labels $\hat{\n}_i$ and
$\eta_i$ that appear as free parameters in $P$ and $\Pobs$
become functions of the integration variable --
$\hat{\n}_i = \hat{\x}_i$ and $\eta_i = \eta_0 - \chi_i$ --
and it is this identification that generates mode-mixing and
wide-angle corrections. For a field with no line-of-sight anisotropy and no radial
evolution, $\Plc$ becomes diagonal, reducing to the standard
isotropic power spectrum. The mixed spectrum $\Pkd$ retains its
dependence on $\d$ through the lightcone geometry, but the
wide-angle corrections vanish and $\Pkd \to P(k)$ in the
plane-parallel limit $s/d \to 0$.

\subsection{Relation to angular power spectra}
\label{sec:angular_spectra}

The angular power spectrum $C_\ell(\chi_1,\chi_2)$ is defined on the lightcone for shells at fixed comoving distance~\citep{1994MNRAS.266..219F,1995MNRAS.275..483H,2013JCAP...11..044D,2013PhRvD..88b3502Y,2018JCAP...03..019T,2018JCAP...10..032T},
\begin{equation}
  \xi_{\rm LC}(\chi_1,\chi_2,\nu)
  =
  \sum_{\ell=0}^\infty
  \frac{2\ell+1}{4\pi}\,
  C_\ell(\chi_1,\chi_2)\,
  \mathcal{L}_\ell(\nu),
  \qquad
  \nu=\hat{\bm n}_1\cdot\hat{\bm n}_2 .
  \label{eq:Cell_def}
\end{equation}
Using Eq.~\eqref{eq:xi_from_Pobs} together with the double-Legendre expansion
of the observer spectrum, Eq.~\eqref{eq:double_legendre_def}, the angular
integral over $\hat{\bm k}$ gives
\begin{equation}
  \xi_{\rm LC}(\chi_1,\chi_2,\nu)
  =
  \sum_{\ell=0}^\infty
  \frac{\mathcal{L}_\ell(\nu)}{2\ell+1}
  \int_0^\infty \frac{k^2\diff k}{2\pi^2}\,
  P^{\rm(O)}_{\ell\ell}(k;\chi_1,\chi_2,\nu).
  \label{eq:xi_from_Pdiag_general}
\end{equation}
Thus scalar angular correlations probe only the diagonal observer multipoles
$P^{\rm(O)}_{\ell\ell}$, but in general these may still depend on the opening
angle $\nu$. The angular spectra are therefore obtained by projecting
Eq.~\eqref{eq:xi_from_Pdiag_general} onto Legendre polynomials:
\begin{align}
  C_L(\chi_1,\chi_2)
  &=
  2\pi \int_{-1}^{1}\diff\nu\,
  \mathcal{L}_L(\nu)\,
  \sum_{\ell=0}^\infty
  \frac{\mathcal{L}_\ell(\nu)}{2\ell+1}
  \,\nonumber \\
  &\times \int_0^\infty \frac{k^2\diff k}{2\pi^2}P^{\rm(O)}_{\ell\ell}(k;\chi_1,\chi_2,\nu).
  \label{eq:Cell_general}
\end{align}
For the separable kernels that arise at linear order,
$P^{\rm(O)}_{\ell\ell}(k;\chi_1,\chi_2)$ carries no non-trivial $\nu$
dependence, and Eq.~\eqref{eq:Cell_general} reduces to
\begin{equation}
  C_\ell(\chi_1,\chi_2)
  =
  \frac{4\pi}{(2\ell+1)^2}
  \int_0^\infty \frac{k^2\diff k}{2\pi^2}\,
  P^{\rm(O)}_{\ell\ell}(k;\chi_1,\chi_2).
  \label{eq:Cell_from_Pdiag}
\end{equation}
The factors of $(2\ell+1)$ reflect the convention used in Eq.~\eqref{eq:double_legendre_def}; with $\mathcal K_\ell=(2\ell+1)i^\ell T_\ell$ this reduces to the standard full-sky expression. Thus the familiar tomographic spectra are  angular projections of the observer spectrum. They retain only the part that remains after the angular average over $\hat{\k}$, which is  the diagonal double-Legendre multipoles $P^{\rm(O)}_{\ell\ell}$. In the linear separable case discussed explicitly below, the off-diagonal coefficients are not independent, see Eq.~\eqref{eq:P_l1l2_result} below. In the general case, however, the off-diagonal coefficients $P^{\rm(O)}_{\ell_1\ell_2}$ with $\ell_1\ne\ell_2$ carry independent angular information, and this is not probed by the scalar $C_\ell$ projection.

\subsection{The observer bispectrum and higher-order statistics}
\label{sec:observer_bispectrum}

The same argument that makes the observer power spectrum diagonal applies to higher-order statistics. For the three-point function, statistical homogeneity on the observer hypersurface gives
\bea\label{eq:Bobs_diagonal}
  &\big\< X^{\rm(O)}(\k_1;\chi_1,\hat{\n}_1)\,
         X^{\rm(O)}(\k_2;\chi_2,\hat{\n}_2)\,
         X^{\rm(O)}(\k_3;\chi_3,\hat{\n}_3) \big\>
  =  \\ \nonumber
  &(2\pi)^3\,\delta^{(3)}_{\rm D}(\k_1{+}\k_2{+}\k_3)\,
    \Bobs(\k_1,\k_2,\k_3;\chi_1,\chi_2,\chi_3,\hat{\n}_1,\hat{\n}_2,\hat{\n}_3),
\eea
which defines the observer bispectrum. Thus the external observer modes obey the usual triangle condition. The lightcone coordinates remain as external labels, just as for the observer power spectrum.

For a local observable, each OFT leg factorises into a source phase and a spatial Fourier mode,
\bea
  &X^{\rm(O)}(\k_i;\chi_i,\hat{\n}_i)
  =
  e^{ik_i\mu_i\chi_i}\,
  X(\k_i,\eta_i;\hat{\n}_i),
  \qquad \\
  &\mu_i=\hat{\k}_i\cdot\hat{\n}_i,\quad
  \eta_i=\eta_0-\chi_i .
\eea
It follows that
\be\label{eq:Bobs_local}
  \Bobs_{\rm local}
  = e^{i(k_1\mu_1\chi_1+k_2\mu_2\chi_2+k_3\mu_3\chi_3)}\,  %B(\k_1,\k_2,\k_3;\eta_1,\eta_2,\eta_3,\hat{\n}_1,\hat{\n}_2,\hat{\n}_3),
B(\k_i;\eta_i,\hat{\n}_i),
\ee
where $B$ is the unequal-time spatial bispectrum of the local field. Using $\k_1+\k_2+\k_3=0$, the phase may equivalently be written as $e^{i(\k_1\cdot\bm{s}_{13}+\k_2\cdot\bm{s}_{23})}$, with $\bm{s}_{i3}=\chi_i\hat{\n}_i-\chi_3\hat{\n}_3$. In the equal-time plane-parallel limit the source separations vanish in this phase, and $\Bobs_{\rm local}$ reduces to the usual spatial bispectrum.

For an integrated observable, each external leg is replaced by its own line-of-sight kernel. For example, if
\be
  X^{\rm(O)}(\k_i;\chi_i,\hat{\n}_i)
  =
  \mathcal I_i(\k_i;\chi_i,\hat{\n}_i)\,\Phi_W(\k_i),
\ee
then
\be\label{eq:Bobs_integrated}
  \Bobs_{\rm int}
  =
  \mathcal I_1(\k_1;\chi_1,\hat{\n}_1)\,
  \mathcal I_2(\k_2;\chi_2,\hat{\n}_2)\,
  \mathcal I_3(\k_3;\chi_3,\hat{\n}_3)\,
  B_\Phi(k_1,k_2,k_3).
\ee
Mixed local--integrated bispectra are obtained by assigning the appropriate phase or line-of-sight kernel to each leg. At higher order, products of local and integrated terms generate the usual internal momentum convolutions, just as in any nonlinear Fourier-space calculation. These nonlinearities change the kernels or vertices of the observable, but not the external diagonality of the observer polyspectra: statistical homogeneity still supplies a single overall delta function enforcing closure of the external wavevectors.

The same reasoning extends to arbitrary $N$-point functions:
\bea
  &\big\< X^{\rm(O)}(\k_1;1)\cdots
         X^{\rm(O)}(\k_N;N) \big\>
  =
  (2\pi)^3\delta^{(3)}_{\rm D}
  \!\left(\sum_{i=1}^N\k_i\right)\\ \nonumber
  &\times S_N^{\rm(O)}(\k_1,\ldots,\k_N;1,\ldots,N).
\eea
For local fields, $S_N^{\rm(O)}$ is the ordinary unequal-time spatial polyspectrum multiplied by one phase per leg; for integrated fields, the corresponding legs carry line-of-sight kernels. The single-observer lightcone $N$-point function is obtained by integrating over the external observer momenta, in the same way as for the two-point function.

\section{Application: Relativistic Number Counts}
\label{sec:application}

%\cc{once you're done look at the overal balance of the section, make sure it tells a story and delivers a set of key points.}

%We now apply the Observer Spectrum framework to the observed galaxy number count fluctuation, $\Delta$. This observable includes not only the standard density and redshift-space distortion (RSD) signals but also relativistic corrections arising from the Doppler effect, gravitational lensing, and local potentials.
%\pritha{We implement the observer spectrum formalism for the observable galaxy number count fluctuation, $\Delta$. 
A key application of the observer spectrum is for the the galaxy number count fluctuation, $\Delta$ which we now consider in detail. In a fully relativistic treatment, the signal contains not only the standard density and redshift-space distortion contributions, but also additional lightcone corrections sourced by Doppler effects and lensing magnification. These terms arise because galaxies are observed on the past lightcone and therefore have contributions both from local source effects and integrated projection effects along the line of sight~\citep{2009PhRvD..80h3514Y,2010PhRvD..82h3508Y,2011PhRvD..84f3505B,2011PhRvD..84d3516C,2012PhRvD..85b3504J,2013JCAP...11..044D,2014JCAP...09..037B,2014JCAP...11..013B,2014PhRvD..90b3513Y,2015CQGra..32s5011B,2021JCAP...12..009M,2026JCAP...06..039A}.

\subsection{
%The Modified Kaiser Kernel 
{The modified kernels}}
\label{sec:delta_kernel}
%We work in the Poisson gauge with the line element $ds^2 = a^2(\eta)[-(1+2\Psi)d\eta^2 + (1-2\Phi)d\x^2]$. Assuming a $\Lambda$CDM cosmology with negligible anisotropic stress ($\Phi=\Psi$), the linear number count fluctuation $\Delta(\x_0; \chi, \hat{\n})$ observed at redshift $z(\chi)$ in direction $\hat{\n}$ is given by \cite{Bonvin:2011bg,Challinor:2011bk}:
\pritha{We work in the Poisson gauge where the line element takes the form %\cc{grown up d's} 
$\mathrm{d}s^2 = a^2(\eta)[-(1+2\Psi)\mathrm{d}\eta^2 + (1-2\Phi)\mathrm{d}\x^2]$. For a $\Lambda$CDM cosmology with negligible anisotropic stress ($\Phi=\Psi$), the galaxy number count fluctuation $\Delta(\x_0; \chi, \hat{\n})$ observed at redshift $z(\chi)$ in direction $\hat{\n}$ can be decomposed as \citep{2011PhRvD..84f3505B,2011PhRvD..84d3516C}:
\begin{equation} \label{numconts}
    \Delta = \Delta_{\rm Newt} + \Delta_{\rm Dop} + \Delta_{\rm Int} 
    %+ \Delta_{\rm Pot} 
    \,.
\end{equation}
Here, $\Delta_{\mathrm{Newt}}$ contains the standard density and Kaiser redshift-space distortion terms. The Doppler contribution, $\Delta_{\mathrm{Dop}}$, arises from peculiar velocities along the line of sight. The integrated contributions, $\Delta_{\mathrm{Int}}$, are the effects along the photon path between the source and the observer, which include %\cc{aren't these all of them?} 
lensing magnification, integrated Sachs-Wolfe contribution and time-delay effects.}
%In the standard Fourier picture, these terms are complicated by the mixing of positions $\x$ and lightcone integrals. In the Observer Spectrum formalism, we relate the Observer Fourier Transform of the number counts, $\Delta^{\rm(O)}(\k; \chi, \hat{\n})$, directly to the primordial matter density fluctuation $\delta_0(\k)$ via a linear transfer function.
\pritha{In the standard Fourier space description, the presence of these light-cone integrals lead to a non-trivial mixing of spatial coordinates $\boldsymbol{x}$, particularly for contributions integrated along the line of sight \citep{1999ApJ...517....1Y,1999ApJ...527..488Y,2022JCAP...01..061C}. With in the observer spectrum formalism, however, we express the observer Fourier transform of the number counts $\Delta^{\rm(O)}(\k; \chi, \hat{\n})$, directly in terms of the primordial matter density fluctuation, $\delta_0(\k)$, through a linear kernel.
We define the modified 
%Kaiser
kernel $\mathcal{K}_\Delta(k, \chi, \mu)$, where $\mu = \hat{\k}\cdot\hat{\n}$, such that:
\begin{equation}
    \Delta^{\rm(O)}(\k; \chi, \hat{\n}) = \mathcal{K}_\Delta(k, \chi, \mu) \, \delta_0(\k).
\end{equation}}
\pritha{The observer spectrum is then simply
\begin{equation}
    \Pobs_\Delta(\k; \chi_1, \chi_2, \hat{\n}_1, \hat{\n}_2) = \mathcal{K}_\Delta(k, \chi_1, \mu_1) \, \mathcal{K}_\Delta^*(k, \chi_2, \mu_2) \, P_0(k),
\end{equation}
where $P_0(k)$ is the 
%primordial 
matter density power spectrum. 
%The kernel $\mathcal{K}_\Delta$ is a sum of four physical components, which we explicitly derive below.
The kernel $\mathcal{K}_\Delta$ is the Fourier space counterpart of (\ref{numconts}) and decomposes into four components. Their explicit forms are derived below.}

\subsubsection{Newtonian Terms}
%The standard "Newtonian" contribution consists of the galaxy density fluctuation (biased by $b$) and the Kaiser RSD term arising from the gradient of the peculiar velocity along the line of sight.
%In real space, $\Delta_{\rm Newt} = [b\delta - \mathcal{H}^{-1}\partial_r(\vvel\cdot\hat{\n})]$. Applying the OFt (Eq.~\ref{eq:OFT_def}), the spatial position $\x_0+\chi\hat{\n}$ yields a phase factor $e^{i k_\parallel \chi}$ (where $k_\parallel = k\mu$). The derivative $\partial_r$ acting on the plane wave pulls down a factor of $ik\mu$.
\pritha{The standard Newtonian contribution  \citep{1987MNRAS.227....1K} consists of the biased density term together with the Kaiser redshift space distortions. The latter arises from the line of sight gradient of the peculiar velocity, $\boldsymbol{v}$. In configuration space, this contribution can be written as 
\begin{equation}
    \Delta_{\rm Newt} = [b\delta - \mathcal{H}^{-1}\partial_{\chi}(\vvel\cdot\hat{\n})]\,.
\end{equation} 
%\cc{define stuff}
where $b$ is the linear bias, $\mathcal{H}$ is the comoving Hubble parameter and $\chi$ is the comoving distance. Applying the observer Fourier transform, defined in Eq.~\ref{eq:OFT_def}, the source position $\x_0+\chi\hat{\n}$ gives rise to the radial phase factor $e^{i k\mu \chi}$. %\cc{stop defining mu} 
The corresponding Newtonian kernel is therefore
\begin{equation}
    \mathcal{K}_{\rm Newt}(k, \chi, \mu) = e^{i k \mu \chi} \, D(\eta) \left[ b(\chi) + f(\chi)\mu^2 \right]\,.
    \label{eq:K_newt}
\end{equation}
Here, $D(\eta)$ is the linear growth factor and $f = \mathrm{d}\ln D/\mathrm{d}\ln a$ is the growth rate.} %The phase factor $e^{i k \mu \chi}$ captures \cc{??} the radial location of the source along the past lightcone.}

\subsubsection{Doppler Terms}
%The Doppler contribution arises from the peculiar velocity of the source and the observer projected along the line of sight. It includes contributions from the standard Doppler effect and the relativistic Euler equation terms.
\pritha{The Doppler contribution arises from the peculiar velocity of the source projected along the line of sight. In real space, this contribution is given by~\citep{2011PhRvD..84f3505B,2011PhRvD..84d3516C,2014PhRvD..89h3535B,2012JCAP...10..025B,2024JCAP...08..027J,Noorikuhani:2022bwc,2021JCAP...12..009M}
\begin{equation}
    \Delta_{\rm Dop} = %- \left( \frac{\mathcal{H}'}{\mathcal{H}^2} + \frac{2-5s}{\mathcal{H}\chi} + 5s - b_e \right) 
    \left(b_e -2\mathcal{Q} + \frac{2(\mathcal{Q} -1)}{\mathcal{H}\chi} - \frac{\mathcal{H}'}{\mathcal{H}^2}\right)(\vvel \cdot \hat{\n})\,,
\end{equation}
where $b_e=\partial(a^3 \bar{n}_g)/\partial\,\mathrm{ln}\,a$ denotes the evolution bias and describes the evolution of the comoving galaxy number density. The magnification bias parameter is defined as $\mathcal{Q}= -\,\partial\,\mathrm{ln}\,\bar{n}_g/\partial\,\mathrm{ln}\,L$, where $L$ is %\cc{don't use 'with $L$ being', rather 'where $L$ is the...'} 
the luminosity threshold of the survey~\citep{2021JCAP...12..009M}. Throughout we assume a $\Lambda$CDM background, for which $\mathcal{H}'/\mathcal{\mathcal{H}}^2 = 1 - 3/2\,\Omega_m$. In Fourier space, the velocity is 
\begin{equation}
    \vvel(\k) = i \frac{\bm k}{k^2} \mathcal{H} f\delta(\k)\,.
\end{equation} 
%\cc{is the sign correct?}
The Doppler contribution to the observer-spectrum kernel becomes}
\pritha{\begin{equation}
    \mathcal{K}_{\rm Dop}(k, \chi, \mu) = ie^{i k \mu \chi} \, D(\eta) \,  \mu f \left(b_e -2\mathcal{Q} + \frac{2(\mathcal{Q} -1)}{\mathcal{H}\chi} - \frac{\mathcal{H}'}{\mathcal{H}^2}\right)  \left(\frac{\mathcal{H}}{k}\right)  \,.
    %- \left[ i \mu \left(\frac{\mathcal{H}_0}{k}\right) \mathcal{A}_{\rm Dop}(0) \right].
    \label{eq:K_dop}
\end{equation}}

%\cc{this is in the wrong place! you haven't defined lensing yet :-)} 

\subsubsection{Lensing Magnification}
\label{sec:lensing_detail}
%The lensing magnification contribution to the number counts is proportional to the convergence $\kappa$. This is an integrated effect, depending on the gravitational potential along the entire line of sight from the observer to the source.
The lensing contribution to the observed galaxy number density contrast is given by 
\begin{equation}
    \Delta_{\rm Lens}(\x_0; \chi, \hat{\n}) = 2(\mathcal{Q}-1) \, \kappa(\x_0; \chi, \hat{\n}).
\end{equation}
The lensing convergence, $\kappa$, quantifies the distorted size of observed images due to gravitational light deflection along the line of sight.  
%\cc{check precision of your language..} 
It is defined as the weighted line-of-sight integral over the transverse Laplacian on the $2$-sphere, $\nabla^2_{\perp}$. 
%\cc{language -- this is the normal potential}.  
Therefore, the relativistic lensing kernel is defined as~\citep{2001PhR...340..291B,2011PhRvD..84f3505B,2011PhRvD..84d3516C}
\begin{align}\label{kappa}
    \kappa(\x_0; \chi, \hat{\n}) &= \frac{1}{2} \int_0^\chi \diff\chi' \, \frac{(\chi-\chi')\chi'}{\chi} \, \nabla_{\perp}^2 \big[ \Phi(\x_0+\chi'\hat{\n}, \eta_0-\chi') \nonumber\\ 
    &+ \Psi(\x_0+\chi'\hat{\n}, \eta_0-\chi')\big] \,.
\end{align}
In the absence of anisotropic stress, the two scalar metric potentials are equal, $\Phi=\Psi$. In addition to the lensing contribution, the observed number count fluctuation also has other integrated line of sight contributions. The kernel can be decomposed into three terms, 
\begin{align}
    \mathcal{K}_{\mathrm{int}}(k,\chi,\mu) = \mathcal{K}_{\mathrm{L}}(k,\chi,\mu) + \mathcal{K}_{\mathrm{TD}}(k,\chi,\mu) + \mathcal{K}_{\mathrm{ISW}}(k,\chi,\mu)\,,
\end{align}
corresponding  to lensing magnification, time delay and the integrated Sachs-Wolfe contribution respectively. 
In the observer Fourier transform, the gravitational potential is written as in Eq.~\ref{lensexp}. Since the potential is evaluated at the position $\x_0+\chi'\hat{\n}$, it picks up the additional phase factor $e^{i k \mu \chi'}$. The transverse Laplacian operator $\nabla_{\perp}^2$ acts on the angular dependence of the field and in the observer Fourier transform, this angular dependence is entirely carried by the phase factor $e^{i k \mu \chi'}$.
\pritha{Substituting the Poisson equation and expressing the density contrast in terms of the linear growth factor, $\delta(\k, \eta) = D(\eta)\delta_0(\k)$, we obtain the full lensing kernel $\mathcal{K}_{\rm Lens}$, %   \cc{are your signs consistent??}
\begin{align}
    \mathcal{K}_{\rm L}(k, \chi, \mu) &= 3(\mathcal{Q}-1) \int_0^\chi \diff\chi' \, \frac{\chi'(\chi-\chi')}{\chi} \, \Omega_m(\chi')\mathcal{H}^2(\chi')\nonumber \\ 
    &D(\chi')\Big[1 - \mu^2 + 2i \frac{\mu}{k\chi'} \Big]e^{i k \mu \chi'}\,.
\end{align}
The remaining integrated contributions originate from the Shapiro time delay term and the integrated Sachs-Wolfe term in the observed galaxy number counts. These terms are suppressed by a factor of $(\mathcal{H}/k)^2$, and therefore are only  important on large scales. Their corresponding kernels are \citep{Noorikuhani:2022bwc,Guedezounme:2024pbj,2011PhRvD..84f3505B,2011PhRvD..84d3516C,2013JCAP...11..044D} %\cc{refs??}
\begin{align}
    \mathcal{K}_{\rm TD}(k, \chi, \mu) &= 6(\mathcal{Q}-1) \int_0^\chi \diff\chi' \, \frac{\Omega_m(\chi')\mathcal{H}^2(\chi')D(\chi')}{k^2\chi}e^{i k \mu \chi'} \,, \\
    \mathcal{K}_{\rm ISW}(k, \chi, \mu) &= 3\Bigg[b_e - 2\mathcal{Q} + \frac{2(\mathcal{Q}-1)}{\chi\mathcal{H}}-\frac{\mathcal{H}'}{\mathcal{H}^2}\Bigg]\nonumber \\
    &\times \int_0^\chi \diff\chi' \, \frac{\Omega_m(\chi')\mathcal{H}^3(\chi')D(\chi')(f(\chi')-1)}{k^2}e^{i k \mu \chi'} \,.
\end{align}

\begin{figure*}
    \centering
    \includegraphics[width=0.9\linewidth]{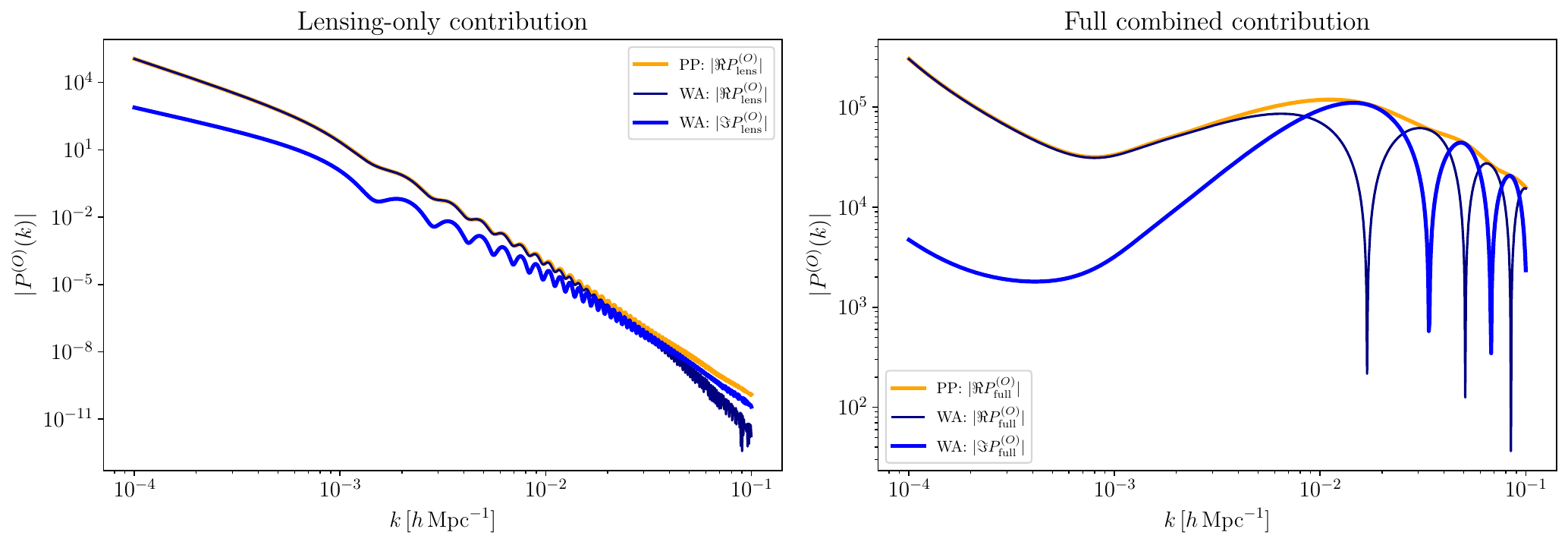}
    \caption{Comparison between the integrated contribution alone (left panel) and the full combined (right panel) to the observed spectrum $P^{(O)}(k)$, shown for both the plane-parallel (PP) and wide-angle (WA) configurations. In the integrated-only case, both the real and imaginary parts are small and decrease with increasing $k$, with the wide-angle result lying below the plane-parallel one over most of the range shown. The plane-parallel result gives the closest comparison with Fig.~2, in \citet{2026JCAP...06..039A}. Qualitatively, both figures show that the lensing and integrated terms are important: although they are smaller than the dominant local contribution on their own, they still affect the shape of the full observed spectrum once all terms are combined.}
    \label{fig:PO_full_with_lensing}
\end{figure*}

\pritha{In Fig.~\ref{fig:PO_full_with_lensing} we compare the observer power spectrum in the plane-parallel and wide-angle configurations. In the plane-parallel case, where the two legs are evaluated at the same redshift and along the same line of sight, the kernel on one leg is identical to that on the other, so the observer spectrum is real. This changes in the wide-angle configuration shown in the right panel. When the two legs are evaluated at different redshift and along different line of sight directions, the two kernels are no longer identical. As a result, the antisymmetric contribution does not cancel, generating a non-vanishing imaginary part of the observer spectrum. This demonstrates that the imaginary component in a genuinely wide angle effect.}

\begin{figure*}
    \centering
    \includegraphics[width=0.95\linewidth]{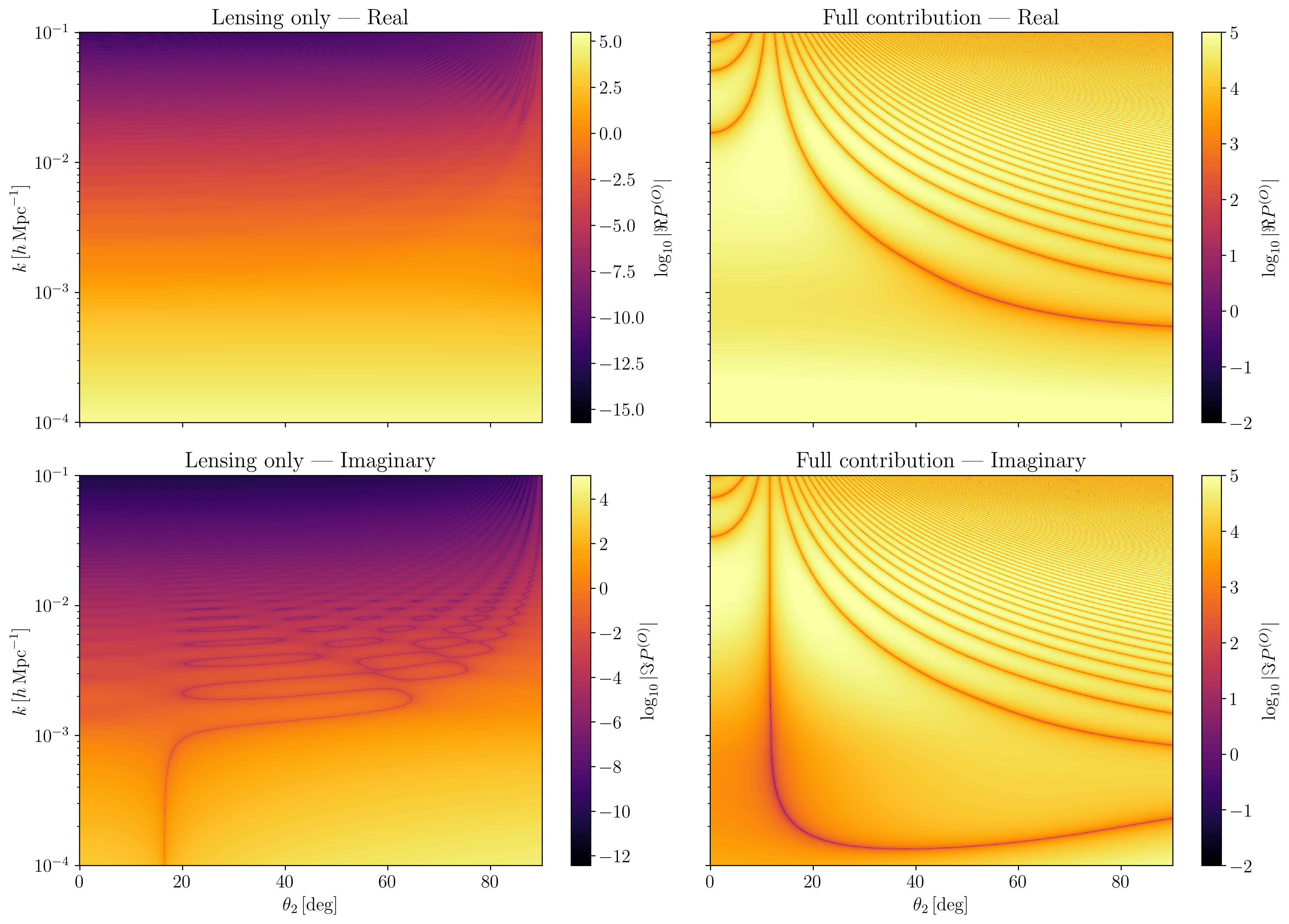}
    \caption{%\cc{Re-do with high resolution -- fonts should be vector format..!} 
    Wide-angle $2$D maps of the observed power spectrum as a function of $k$ and $\theta_2$ for the configuration $z_1 = 1.5, z_2 = 1.55$, and $\theta_1 = 0$. The left panels show the lensing only contribution, while the right panels show the full contribution including local and integrated relativistic terms. The top row shows the real part, $\Re P^{(O)}$, and the bottom row shows the imaginary part, $\Im P^{(O)}$. The full contribution is large in amplitude and shows a rich angular structure, while a notable feature of the lensing contribution is its distintive angular structure in the $(k,\theta_2)$ plane. }
    \label{fig:2D-wideangle}
\end{figure*}
Fig.~\ref{fig:2D-wideangle} presents the wide-angle power spectrum as a function of $k$ and $\theta_2$ for $z_1 = 1.5$, $z_2 = 1.55$, and $\theta_1 =0$, showing separately the lensing-only contribution and the full result. The lensing contribution is characterised by a curved structure in the $(k,\theta_2)$ plane. }

\subsubsection{Legendre Decomposition of the Kernel}
\label{sec:kernel_legendre}
%\cc{is this bit too long? I think you can merge the local vs integrated terms.}
%The modified Kaiser kernel $\mathcal{K}_\Delta(k, \chi, \mu)$ is a complex-valued function of the angle $\mu$ between the wavevector and the line of sight. To facilitate comparison with wide-angle and spherical formalisms, we decompose this angular dependence into Legendre multipoles:
\pritha{The modified Kaiser kernel depends on $\mu$, the cosine of the angle between the wavevector and the line of sight. In order to compare directly with wide angle treatments and spherical analysis, it is useful to decompose its angular dependence into Legendre multipoles.
\begin{equation}
    \mathcal{K}_\Delta(k, \chi, \mu) = \sum_{\ell=0}^\infty \mathcal{K}_\ell(k, \chi) \, \mathcal{L}_\ell(\mu).
\end{equation}
%The multipole coefficients are obtained by projecting the kernel onto the Legendre basis. 
%\begin{equation}
%    \mathcal{K}_\ell(k, \chi) = \frac{2\ell+1}{2} \int_{-1}^{1} \diff\mu \, \mathcal{K}_\Delta(k, \chi, \mu) \, \mathcal{L}_\ell(\mu)\,.
%\end{equation}
Since the kernel contains the oscillatory phase, $e^{i k \mu \chi}$ (or $e^{i k \mu \chi'}$ for lensing), these projections can be naturally expressed in terms of spherical Bessel functions $j_{\ell}(y)$, where $y=k\chi$.} %This expansion converts the angular dependence of the Fourier-space kernel into a spherical multipole basis, which is the natural framework for wide angle and spherical analysis.} 

\paragraph{Newtonian Terms:}

The Newtonian contribution in Eq.~\ref{eq:K_newt} can be expanded in a spherical multipole basis. Using the plane-wave expansion, we have 
\begin{equation}
    e^{iy\mu}
    =
    \sum_{\ell=0}^{\infty}
    (2\ell+1)i^\ell j_\ell(y)\mathcal{L}_\ell(\mu)\,.
\end{equation}
The density term gives the \(j_\ell(y)\) contribution, while the RSD term can be written in terms of derivatives of the radial phase. Writing this in the Legendre basis then gives
\begin{equation}
    \mathcal{K}_\ell^{\rm Newt}(k,\chi) = (2\ell+1)i^\ell D(\eta)
    \left[b\,j_\ell(y) - f\,j_\ell''(y)\right].
    \label{eq:Kell_newt}
\end{equation}
Here primes denote derivatives with respect to the argument \(y\). Keeping the full radial phase \(e^{ik\mu\chi}\) changes the usual Fourier-space Kaiser angular dependence into a spherical multipole expansion. In the plane-parallel limit, the Kaiser term has a simpler angular dependence. This means it can be written using a few Legendre multipoles, rather than infinitely many. However, once the full radial phase is kept, the kernel contributes to all spherical multipoles \(\ell\). 

\paragraph{Doppler Terms:}
%The Doppler term (Eq.~\ref{eq:K_dop}) contains a factor of $\mu e^{i y \mu}$, which corresponds to a derivative with respect to the argument. Its projection yields:
The Doppler contribution (\ref{eq:K_dop})carries a factor of $\mu e^{i y \mu}$, as it has an odd dependence on the line of sight. Therefore, it has the first derivative with respect to the argument. Its multipole decomposition takes the form 
\begin{align}
    \mathcal{K}_\ell^{\rm Dop}(k, \chi) &= (2\ell+1) i^\ell \Big[ D(\eta) \frac{\mathcal{H}(\chi)}{k} \nonumber\\ 
    &\left(b_e -2\mathcal{Q} + \frac{2(\mathcal{Q} -1)}{\mathcal{H}\chi} - \frac{\mathcal{H}'}{\mathcal{H}^2}\right)j_\ell'(k\chi)\Big]\,. 
\end{align}
   % The subtraction term ensures that the kernel vanishes as $\chi \to 0$ (where $j_1'(0) = 1/3$), enforcing the condition that the observer dipole is included in this kernel. [Taken the source terms out for now]
%\begin{figure}[h]
%    \centering
%    \includegraphics[width=0.78\linewidth]{doppler_diagonal_multipoles.png}
 %   \caption{Caption}
%    \label{fig:placeholder}
%\end{figure}
\paragraph{Lensing Terms:}
Unlike the local density and RSD terms, the lensing term is integrated along the line of sight and therefore depends on the phase evaluated at $\chi'$. Starting from the Fourier space lensing kernel, the angular dependence appears through the factor $[1-\mu^2 + 2i\mu/(k\chi')]e^{ik\mu \chi'}$. When projected onto the Legendre multipoles, this gives a combination of spherical bessel functions and their derivatives. % The transverse Laplacian acts only on the angular dependence of the mode 
%\begin{equation}
%    \nabla_{\perp}^2[j_{\ell}(k\chi)Y_{\ell m}(\hat{\boldsymbol{n}})] = - \frac{\ell(\ell+1)}{\chi^2} j_{\ell}
%\end{equation}
%\begin{equation}
%    \nabla_\perp^2
%    =
%    \frac{1}{\chi'^2}\nabla_\Omega^2 .
%\end{equation}
%\begin{equation}
%    \nabla_{\hat{\bm{n}}}^2 Y_{\ell m}(\mathbf{n})
%=
%    -\ell(\ell+1)Y_{\ell m}(\mathbf{n})\,.
%\end{equation}
Using the spherical Bessel equation below, the derivative combination can be simplified, leaving the projected lensing term proportional to $j_\ell$ alone,
%Equivalently, this result follows from the spherical Bessel differential equation,\cc{why chi-prime? - changed to $\chi$}
\begin{equation}
    j''_{\ell}(k\chi)+ \frac{2}{k\chi}j'_{\ell}(k\chi) +
    \left[1 - \frac{\ell(\ell+1)}{(k\chi)^2} \right] j_{\ell}(k\chi) = 0 .
\end{equation}
%This identity shows how the transverse momentum factor in the Cartesian expression is converted into the angular eigenvalue $\ell(\ell+1)$ after written in the spherical basis. Consequently, the lensing multipole kernel is
Substituting this into the line of sight integral gives the lensing multipole kernel, 
\begin{align}
    \mathcal{K}_\ell^{\rm Lens}(k,\chi)
    &= 3(2\ell+1)i^\ell (\mathcal{Q}-1) \frac{\ell(\ell+1)}{k^2}
    \int_0^\chi\diff\chi'\,\frac{\chi'(\chi-\chi')}{\chi} \nonumber \\
    &\quad \times \Omega_m(\chi')\mathcal{H}^2(\chi')D(\eta_0-\chi')\frac{j_\ell(k\chi')}{\chi'^2}\,.
\end{align}
The time-delay and ISW terms are also integrated along the line of sight, but they do not contain the transverse Laplacian. Their projections therefore involve the spherical Bessel function directly, without the additional factor of $\ell(\ell+1)$. The corresponding kernels are
\begin{align}
    \mathcal{K}_\ell^{\rm TD}(k,\chi)
    &=6i^\ell (\mathcal{Q}-1) \frac{(2\ell+1)}{k^2\chi}
  %  \nonumber \\
 %   &\quad \times
    \int_0^\chi \diff\chi'\, \Omega_m(\chi') \mathcal{H}^2(\chi')  \nonumber\\
    &\quad \times D(\eta_0-\chi')j_\ell(k\chi')\,, \\
    \mathcal{K}_\ell^{\rm ISW}(k,\chi)
    &=3i^\ell 
    \left[b_e - 2\mathcal{Q} + \frac{2(\mathcal{Q}-1)}{\chi\mathcal{H}} - \frac{\mathcal{H}'}{\mathcal{H}^2} \right] \frac{(2\ell+1)}{k^2}
    \nonumber \\
    &\quad \times \int_0^\chi \diff\chi'\, \Omega_m(\chi') \mathcal{H}^3(\chi') \left[f(\chi')-1\right] \nonumber\\
    &\quad \times D(\eta_0-\chi') j_\ell(k\chi')\,.
\end{align}

\subsubsection{Relation to standard spherical kernels}
\label{sec:standard_kernel_relation}
\pritha{In the full-sky line-of-sight approach~\citep{1994MNRAS.266..219F,1995MNRAS.275..483H,1998ASSL..231..185H,2013PhRvD..88b3502Y,2013JCAP...11..044D,2018JCAP...03..019T,2018JCAP...10..032T,2024PhRvD.109h3502G,2024arXiv240404812W,2024arXiv240404811B}, %(e.g.\ Heavens, Dor\'e et al.)
a given contribution to the number counts is written as
\begin{equation}
  \Delta(\chi,\hat{\bm{n}})
  = \int \frac{\mathrm{d}^3 k}{(2\pi)^3}
    \sum_{\ell m} 4\pi\,i^\ell\,
    T_\ell(k,\chi)\,
    Y_{\ell m}(\hat{\bm{n}})\,Y_{\ell m}^*(\hat{\bm{k}})\,
    \delta_0(\bm{k}),
  \label{eq:DeltaX_spherical}
\end{equation}
where $T_\ell(k,\chi)$ is the usual full-sky transfer kernel. It already contains the spherical Bessel factor  $j_\ell(k\chi)$, together with any angular dependence from the source. For the observer-spectrum approach, the same contribution is instead written as
\begin{align} \label{eq:DeltaX_OFT}
  &\Delta(\chi,\hat{\bm{n}})
  = \int \frac{\mathrm{d}^3 k}{(2\pi)^3}
    \sum_{\ell} \mathcal{K}_\ell(k,\chi)\,
    \mathcal{L}_\ell(\hat{\bm{k}}\!\cdot\!\hat{\bm{n}})\,
    \delta_0(\bm{k})\, \\
    &\,\mathrm{where}\,\, \\ \nonumber
 & \mathcal{K}(k,\chi,\mu) 
  = \sum_\ell \mathcal{K}_\ell(k,\chi)\,\mathcal{L}_\ell(\mu)\,.
\end{align}
Using the addition theorem
\(
  \mathcal{L}_\ell(\hat{\bm{k}}\!\cdot\!\hat{\bm{n}})
  = \frac{4\pi}{2\ell+1}\sum_m
    Y_{\ell m}(\hat{\bm{n}})\,Y_{\ell m}^*(\hat{\bm{k}})
\),
and comparing \eqref{eq:DeltaX_spherical} with \eqref{eq:DeltaX_OFT} term-by-term, we get the exact relation
%\begin{equation}
  %\boxed{\;
%    \mathcal{K}^X_\ell(k,\chi)
%    = (2\ell+1)\,i^\ell\,T^X_\ell(k,\chi)
%    \quad\Longleftrightarrow\quad
%    T^X_\ell(k,\chi)
%    = i^{-\ell}\,\frac{\mathcal{K}^X_\ell(k,\chi)}{2\ell+1}
  %\;}
%  \label{eq:Kl_Tl_mapping}
%\end{equation}
%\begin{equation}
%    \mathcal{K}^X_\ell(k,\chi)
%    = (2\ell+1)\,i^\ell\,T^X_\ell(k,\chi) .
%    \label{eq:K_from_T_mapping}
%\end{equation}
\begin{equation}
    T_\ell(k,\chi)
    = i^{-\ell}\,\frac{\mathcal{K}_\ell(k,\chi)}{2\ell+1} .
    \label{eq:T_from_K_mapping}
\end{equation}
\paragraph{Local terms.}
For any local contribution, such as the density, Kaiser RSD, Doppler, the kernel is evaluated
at the source position. In this case, the full-sky transfer kernel is related to the observer-spectrum multipole using the same relation as above, Eq.~\eqref{eq:T_from_K_mapping}.}
%\begin{equation}
%  T^{X}_{\ell}(k,\chi)
%  = i^{-\ell}\,\frac{\mathcal{K}^{X}_\ell(k,\chi)}{2\ell+1},
%  \label{eq:local_T_from_K}
%\end{equation}
%so the usual full-sky kernel $T^X_\ell$ is just the OFT multipole divided by $(2\ell+1)i^\ell$.
%All the plane-wave phase $e^{ik\mu\chi}$ has already been converted into spherical Bessel
%functions inside $T^X_\ell$.}

\paragraph{Integrated terms.}
For the integrated contributions, such as lensing and ISW-like terms, the kernel is not evaluated only at the source position. Instead, it is all the contributions along the line of sight~\citep{2013JCAP...11..044D,2018JCAP...03..019T,2018JCAP...10..032T,2011PhRvD..84d3516C}. For a source at distance $\chi_s$, the observer-spectrum multipole can be written as
\begin{equation}
  \mathcal{K}_\ell(k,\chi_s)
  = \int_0^{\chi_s}\!\mathrm{d}\chi\,
    W(\chi_s,\chi)\,
    \mathcal{K}^{\mathrm{loc}}_\ell(k,\chi),
  \label{eq:Kl_integrated}
\end{equation}
and the standard full-sky transfer function is
\begin{equation}
  T_\ell(k,\chi_s)
  = i^{-\ell}\,\frac{\mathcal{K}_\ell(k,\chi_s)}{2\ell+1}
  = \int_0^{\chi_s}\!\mathrm{d}\chi\,
    W(\chi_s,\chi)\,
    T^{\mathrm{loc}}_\ell(k,\chi)\,.
  \label{eq:Tl_integrated}
\end{equation}

\color{black}

\subsubsection{Observer terms}
\label{sec:observer_terms}

A subtle but important point in the  relativistic description of galaxy clustering is the treatment of observer-position contributions, particularly with respect to their effect on the observed statistics. The Doppler effect is the clearest example: the observer's velocity (in the Newtonian gauge) enters the number-count fluctuation not as the formal $\chi \to 0$ limit of the source velocity, but as a distinct boundary term evaluated at the observer's location. In standard relativistic treatments~\citep{2009PhRvD..80h3514Y,2010PhRvD..82h3508Y,2011PhRvD..84f3505B,2011PhRvD..84d3516C,2014CQGra..31w4001Y,2011PhRvD..84f3505B,2011PhRvD..84d3516C,2009PhRvD..80h3514Y}, this boundary term combines with other observer-position contributions to the redshift and volume perturbations.

In the observer spectrum we can schematically split the Doppler kernel into source and observer components:
\begin{equation}
  \mathcal{K}_{\rm Dop}(k,\chi,\mu)
  =
  e^{ik\mu\chi}\,\mathcal{K}_{\rm Dop}^{\rm src}(k,\chi,\mu)
  +
  \mathcal{K}_{\rm Dop}^{\rm obs}(k,\mu)\,.
\end{equation}
The key difference is that $\mathcal{K}_{\rm Dop}^{\rm obs}$ has no radial phase $e^{ik\mu\chi}$, since the observer is located at the origin of the integration variable $\x_0$. Consequently, the observer term represents a coherent boundary mode rather than an oscillatory field.

The handling of these terms depend on the context. For an ensemble of observers in an $N$-body simulation say, the observer velocity is a stochastic quantity. In this case, it needs to be treated on the same footing as the source terms. 
In an actual galaxy survey, however, observations are made from a single, fixed realization of the observer's velocity. Because it lacks the oscillatory radial phase, this single realization acts as a coherent global dipole (or monopole for other observer terms such as the potential) across the sky, rather than as a fluctuating clustering signal. In standard data analysis, these fixed observer-position terms are either absorbed into the definition of the observed mean number density or  projected out by the estimator via monopole and dipole subtraction. For practical power-spectrum analyses one may therefore set
\begin{equation}
  \mathcal{K}_{\rm Dop}^{\rm obs}=0\,.
\end{equation}
Various perspectives on this subtle issue are discussed in~\citet{2020JCAP...11..064G,2021MNRAS.504.5612D,2022JCAP...01..061C}.

\subsection{The Angular Structure of the Observer Spectrum}
\label{sec:angular_structure}

\pritha{The Observer Spectrum $\Pobs_\Delta(\k; \chi_1, \chi_2, \hat{\n}_1, \hat{\n}_2)$ depends on the direction of the wavevector $\k$ only through the angles with the two lines of sight. Hence, we write the angular dependence as a double Legendre expansion. This separates the physical contributions from the wide-angle geometry and avoids using perturbative expansions for the RSD terms. The angular dependence is instead kept through the Legendre multipoles. 

\subsubsection{The Double-Legendre Expansion}
We define the multipole coefficients $P_{\ell_1 \ell_2}(k; \chi_1, \chi_2)$ via the expansion given in Eq.~\ref{eq:double_legendre_def}. %\cc{this was defined earlier}
%\begin{equation}
%    \Pobs_\Delta(k; \chi_1, \chi_2, \mu_1, \mu_2) = \sum_{\ell_1=0}^\infty \sum_{\ell_2=0}^\infty P_{\ell_1 \ell_2}(k; \chi_1, \chi_2) \, \mathcal{L}_{\ell_1}(\mu_1) \, \mathcal{L}_{\ell_2}(\mu_2)\,.
%    \label{eq:double_legendre_def}
%\end{equation}
Using the orthogonality relation for the Legendre polynomials, we can write the coefficients as 
\begin{align}
    P_{\ell_1 \ell_2}(k; \chi_1, \chi_2) =& \frac{(2\ell_1+1)(2\ell_2+1)}{4} \int_{-1}^{1} \diff\mu_1 \int_{-1}^{1} \diff\mu_2 \, \Pobs_\Delta(\k) \, \nonumber\\
    &\quad \times \mathcal{L}_{\ell_1}(\mu_1) \mathcal{L}_{\ell_2}(\mu_2)\,.
\end{align}
This decomposition completely separates the angular dependence into the two Legendre factors. The coefficients $P_{\ell_1 \ell_2}$ depend only on the wavenumber $k$ and the two radial distances $\chi_{1,2}$.

\subsubsection{Derivation from Kernel Multipoles}
We can now write the coefficients explicitly using the kernel expansion. Substituting the kernel decomposition in Legendre basis and substituting it into the definition of the observer spectrum gives 
%\begin{equation}
%    \mathcal{K}_\Delta(k,\chi,\mu) = \sum_\ell \mathcal{K}_\ell(k,\chi) \mathcal{L}_\ell(\mu)
%\end{equation}
%into the definition of the observer spectrum gives 
\begin{align}
    \Pobs_\Delta &= \left[ \sum_{\ell_1} \mathcal{K}_{\ell_1}(k, \chi_1) \mathcal{L}_{\ell_1}(\mu_1) \right] \left[ \sum_{\ell_2} \mathcal{K}_{\ell_2}^*(k, \chi_2) \mathcal{L}_{\ell_2}(\mu_2) \right] P_0(k) \nonumber \\
    &= \sum_{\ell_1, \ell_2} \left[ \mathcal{K}_{\ell_1}(k, \chi_1) \mathcal{K}_{\ell_2}^*(k, \chi_2) P_0(k) \right] \mathcal{L}_{\ell_1}(\mu_1) \mathcal{L}_{\ell_2}(\mu_2)\,.
\end{align}
Comparing this with Eq.~\eqref{eq:double_legendre_def}, we get %the fundamental factorization result:
\begin{equation}
    P_{\ell_1 \ell_2}(k; \chi_1, \chi_2) = \mathcal{K}_{\ell_1}(k, \chi_1) \, \mathcal{K}_{\ell_2}^*(k, \chi_2) \, P_0(k)\,.
    \label{eq:P_l1l2_result}
\end{equation}
%This result is remarkably simple. 
The double-Legendre multipoles are therefore built from the two kernel multipoles derived in Sec.~\ref{sec:double_legendre}, one for each line of sight. 

\begin{figure*}
    \centering
    \includegraphics[width=0.9\linewidth]{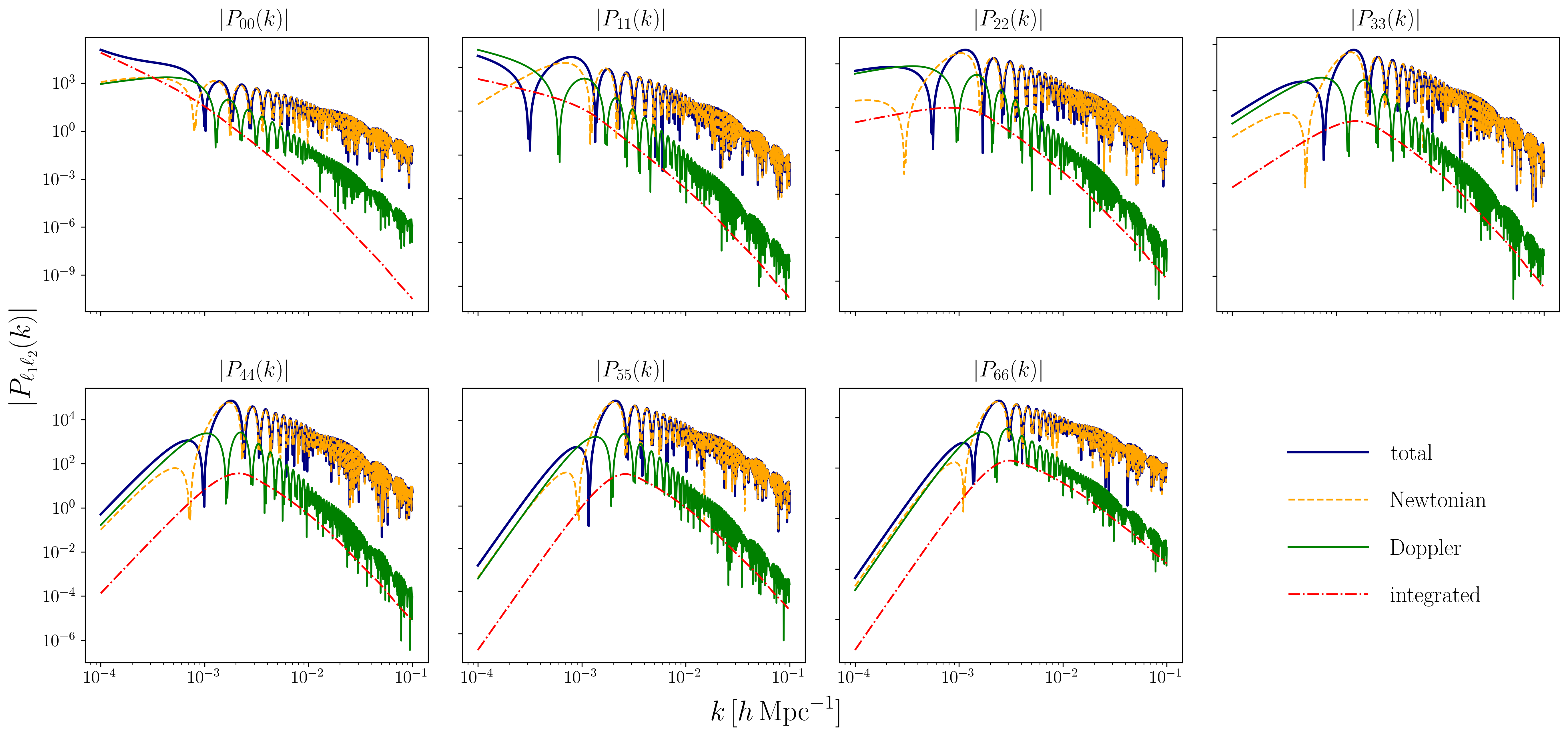}
    \caption{%\cc{Add in P00. either keep 2 rows and add P77 or have 1 row up to P44} 
    Absolute values of the double-Legendre multipoles $|P_{\ell\ell}|$ for $\ell = 1,...,6$. The total observed contribution is shown in blue, with the Newtonian, Doppler and integrated contributions shown separately in orange dashed, green and red dot-dashed lines, respectively. Across all multipoles, the Newtonian term provides the dominant contribution over a wide range of scales, particularly towards intermediate and smaller scales. On the largest scales, however, the relativistic Doppler and integrated terms become more visible and contribute to the total signal. The relative size and oscillatory behaviour of these contributions varies with $\ell$, showing that the angular structure of the observed spectrum is sensitive to the different terms entering the relativistic kernel.}
    %Comparison of the absolute values of the diagonal multipoles $|P_{\ell\ell}|$ for $\ell = 1,...,6$, showing the total contribution (blue), the Newtonian term (orange dashed), the Doppler term (green), and the integrated terms (red dot-dashed) as functions of wavenumber $k$. The total signal is dominated by the Newtonian contribution over most scales, while Doppler and integrated effects become more relevant on the largest scales, with their relative importance varying across multipoles. }
    \label{fig:multipoles}
\end{figure*}

%\cc{the descriptions are good. is it possible to relate to the same effects in 'normal' power spectra, maybe refer to addis' plots?} 
Fig.~\ref{fig:multipoles} shows the diagonal multipoles $P_{\ell\ell}(k)$ for $\ell = 0$ to $6$, decomposed into their total, Newtonian, Doppler and integrated contributions. We see that the Newtonian term provides the dominant contribution over most of the $k$-range, and largely determines the overall shape of the total signal, especially on intermediate and smaller scales. The Doppler and integrated terms are subdominant in amplitude, but their relative importance increases towards large scales. Their impact also depends on the multipole considered: for some $\ell$, the Doppler contribution becomes comparable to the Newtonian part over a range of scales. The integrated term remains smoother and generally smaller, though still non-negligible on the largest scales. }

\begin{figure*}
    \centering
    \includegraphics[width=0.9\linewidth]{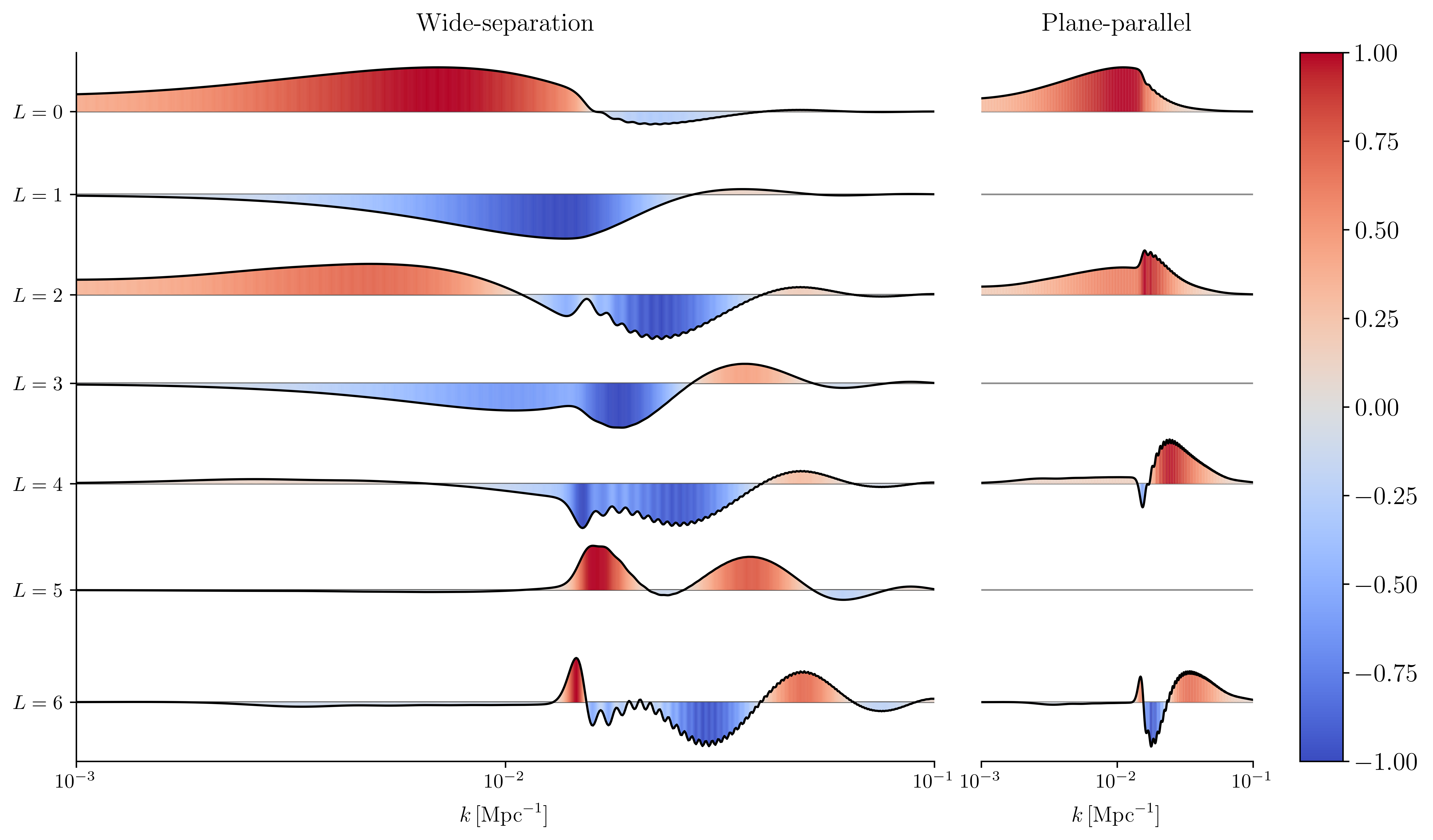}
    \caption{
    %Plot of the multipoles $P_L(k)$ as a function of $k$, shown here for $L=0$ to $L=6$, for the configuration $z_1=1$ and $z_2 = 1.05$. The left panel shows the full observer-spectrum result as a function of wavenumber $k$, while the right panel shows the corresponding plane-parallel (PP) limit. In the full result, both even and odd multipoles are present, with even multipoles purely real and odd multipoles purely imaginary. In the plane-parallel limit, only the even multipoles remain, while the odd multipoles vanish. The colour scale indicates the sign of the underlying signal, with red denoting positive values and blue denoting negative values. Lower-order multipoles are dominated by broader positive features, while higher order multipoles exhibit increasingly oscillatory behaviour.
    {
    Plot of the normalised plane parallel multipoles $P_L(k)$ as a function of $k$, shown here for $L=0$ to $L=6$, for the separation $z_1=1$ and $z_2 = 1.05$ (left) compared to the plane parallel equal redshift case $z_1=z_2=1$ (right). 
%    The left panel shows the full observer-spectrum result as a function of wavenumber $k$, while the right panel shows the corresponding plane-parallel (PP) limit. 
    In the full result, both even and odd multipoles are present, with even multipoles purely real and odd multipoles purely imaginary. In the plane-parallel limit the odd multipoles vanish, but the  separation in redshift breaks the line-of-sight symmetry giving rise to odd multipoles. The colour scale indicates the sign of the underlying signal, with red denoting positive values and blue denoting negative values. Lower-order multipoles are dominated by broader positive features, while higher order multipoles exhibit increasingly oscillatory behaviour.}}
    \label{fig:P_L_plot}
\end{figure*}
Fig.~\ref{fig:P_L_plot} shows the multipoles $P_L(k)$ for the full observed power spectrum. The low-order multipoles are broader and smoother. As $L$ increases, the structure becomes increasingly oscillatory, with sharper peaks and more frequent sign changes. A clear difference is also visible between the even and odd multipoles. The even multipoles show more pronounced oscillatory behaviour, while the odd multipoles are comparatively smoother across the same range of scales. This configuration is plane-parallel like, since the two directions are aligned, $\theta_1 = \theta_2$. However, the redshifts are different $z_1 \neq z_2$, so the signal still carries a non-trivial dependence on the radial separation between the two points. In the strict plane-parallel limit, only the even multipoles survive, while the odd multipoles vanish. The resulting structure should therefore be understood as arising primarily fom unequal redshift, rather than from a non-zero opening angle. %\cc{good!} 

%\clearpage

\color{black}

\section{Conclusions}
\label{sec:discussion}

We have defined the observer power spectrum $\Pobs$ and other poly-spectra by Fourier
transforming fields over observer positions on a spatial hypersurface at fixed
lightcone coordinates, rather than over source positions on a single
lightcone. Because the integration domain is a spatial slice,
translational invariance is preserved and $\Pobs$ is diagonal in $\k$
for any observable, local or integrated, and wide separation effects are fully included. The lightcone geometry is not integrated over but it is also not
removed: it appears in the external labels
$(\chi_i,\hat{\n}_i)$ rather than as mode-mixing between different
Fourier modes. For local fields, $\Pobs$ is the unequal-time spatial
power spectrum times a  phase $e^{i\k\cdot\bm{s}}$; for
integrated observables such as lensing it factorises into
line-of-sight transfer functions acting on each leg independently,
proportional to $P_\Phi(k)$ at the same $\k$ rather than a convolution
over Fourier modes. The same per-leg structure extends to the observer
bispectrum and to $N$-point statistics at linear perturbative order.

The closest existing object to $\Pobs$ in the literature is the `theory power spectrum' $P^{\rm th}$ of~\citet{2020JCAP...11..064G} \citep[see also][]{2010PhRvD..82h3508Y,2013PhRvD..88b3502Y}, a diagonal Fourier-space two-point function for the observed galaxy fluctuation defined on a hypersurface of simultaneity at the observed redshift, and different from the power spectrum measured on the lightcone. At linear order, for the same observed fluctuation, $\Pobs$ and $P^{\rm th}$ mathematically differ  by the phase convention associated with anchoring the Fourier expansion at the observer rather than at the source. The point of our observer based definition is that this diagonal object is not obtained by relying on transfer functions in linear theory, but by applying the Fourier transform to the variable that  obeys statistical homogeneity. This gives a definition that applies to any lightcone observable, local or integrated, including wide angle and separation effects, and extends directly beyond linear theory to nonlinear observables and $N$-point statistics. The lightcone coordinates $(\chi_i,\hat{\n}_i)$ remain as external labels, making explicit how the usual lightcone statistics are obtained from the same underlying object.

The principal two-point statistics used in large-scale structure
analysis are each given as projections of $\Pobs$~-- see Fig.~\ref{fig:hierarchy}. The angular power spectrum
$C_\ell(\chi_1,\chi_2)$ is the scalar projection of the diagonal
multipoles $P_{\ell\ell}^{\rm(O)}$; the lightcone correlation function
$\xi_{\rm LC}$ is the momentum-space integral of $\Pobs$; the mixed
spectrum $\bar{P}(\k,\d)$ and the non-diagonal kernel
$\tilde{P}(\k_1,\k_2)$ are lightcone projections in which the external
labels become functions of the integration variable, which is the
origin of their mode-mixing and wide-angle corrections. The expectation
value of the Yamamoto estimator is a windowed projection of $\Pobs$
(Eq.~\ref{eq:yamamoto_from_Pobs}), which is exact in both wide-angle and
integrated regimes. 

Treating the observer's position $\x_0$ as the integration variable
also clarifies the status of observer terms such as
$\vvel_o\cdot\hat{\n}$ in the Doppler kernel. At the ensemble level
these are stochastic on the same footing as source terms, but they
lack the radial phase $e^{ik\mu\chi}$ carried by source contributions,
so in a single realisation they act as coherent boundary modes rather
than as oscillatory clustering contributions across the volume. The framework is therefore ideal for
 simulations, where the ensemble of observer positions is
given by the final time-step of the simulation: $\Pobs$ can be measured by correlating
lightcone rays constructed from an ensemble of observers, with no
reference to a survey window, before projecting onto $\xi_{\rm LC}$,
$C_\ell$ or the Yamamoto multipoles. This may give an new simpler method to validate the theory of relativistic effects, for example, which is quite difficult for the normal power spectrum~\citep{2021MNRAS.501.2547G}. 

We have illustrated the formalism with the full linear relativistic
kernel for the observed number count fluctuation, including density,
Kaiser RSD, Doppler, lensing magnification, time-delay and integrated
Sachs-Wolfe contributions, and shown the explicit mapping between its
Legendre multipoles and the transfer functions used in the standard
full-sky $C_\ell$. The Appendix
shows how the usual wide-angle expansion of $\bar{P}(\k,\d)$ in powers
of $s/d$ is recovered from $\Pobs$, separating the intrinsic angular
dependence of the field statistics from the geometric Jacobian and the
Fourier projection of the estimator. A natural next step is to apply
the same construction to the observer bispectrum, for which the
diagonal structure and per-leg factorisation extend directly, and to
validate $\Pobs$ against lightcone measurements in $N$-body
simulations.

The concept of the observer spectrum is quite general. A natural extension is to formulate the construction independently of the flat-FLRW, plane-wave representation used here.  In curved FLRW the observer FT should be replaced by an expansion in the scalar harmonics of the constant-curvature observer hypersurface. In homogeneous but anisotropic backgrounds, such as Bianchi models, the corresponding modes are those adapted to the spatial homogeneity group; the spectrum still diagonalises but is no longer rotationally invariant, acquiring dependence on the orientation of $\k$ relative to the preferred axes. In radially inhomogeneous models such as LTB, or in a general spacetime, the situation is different: without spatial homogeneity there is no global momentum-conserving delta function and no diagonal three-dimensional power spectrum. The fully covariant object is instead a two-observer lightcone 2-point function, defined for a chosen observer congruence $u^a$ as $\Xi[p,q;\lambda_1,\hat \n_1;\lambda_2,\hat \n_2]=\langle X[p;\lambda_1,\hat \n_1]X[q;\lambda_2,\hat \n_2]\rangle$, where $p$ and $q$ are events on the congruence and $(\lambda,\hat \n)$ are the affine parameter along the past null geodesic from each event and its direction in the local rest frame. The affine parameter is not itself an observable, but it is the  natural  coordinate on the past lightcone. Direct observables~-- redshift, angular diameter distance, flux~-- are functions of $(\lambda,\hat n)$, but using them directly in the 2pcf is difficult because they are not monotonic. The observer spectrum developed in this paper is the flat, statistically homogeneous transform of $\Xi$, in which the dependence on the two observer positions reduces to their separation and can therefore be diagonalised by Fourier transform. This more general covariant formulation is left for future work.

\section*{Acknowledgements}
The authors are supported by STFC (UK) grant ST/X000931/1. We thank  Stefano Camera and Roy Maartens for discussions.

\section*{Data Availability}
No new data were generated or analysed in support of this research.

\color{black}
%\clearpage

\appendix

\section{Wide-angle effects: intrinsic versus projection}

The spectrum $P^{(\rm O)}$ encodes the exact two-point statistics of any observable without reference to a plane-parallel limit. Consequently, the angular power spectra $C_\ell$ follows directly from $P^{(\rm O)}$ without requiring any wide-angle expansion.
Nevertheless, standard estimators such as $\bar{P}_L(k,d)$ are defined via Fourier transforms over pair separations, and their relationship to the underlying field statistics is typically treated via a wide-angle expansion in powers of $s/d$~\citep{1998ApJ...498L...1S,2000ApJ...535....1M,2004ApJ...614...51S,2008MNRAS.389..292P,2016JCAP...01..048R,2018MNRAS.476.4403C,2019JCAP...03..040B,2023JCAP...04..067P}, mainly to simplify the analysis. In this appendix, we show how this perturbative expansion is recovered from $\Pobs$. This formulation systematically splits wide-angle corrections into three  components: first, the intrinsic angular dependence of the field statistics; second, the geometric distortion of the lightcone; and finaly  the Fourier-space projection defining the estimator.

The mixed spectrum $\bar{P}(\k,\d)$ is the Fourier transform of the lightcone correlation function over the pair separation $\s$ at fixed midpoint $\d$:
\be\label{eq:Pkd_def_app}
  \bar{P}(\k,\d) = \int \frac{\diff^3 k'}{(2\pi)^3} \int \diff^3 s\;e^{-i\k\cdot\s}\; \Pobs\!\left(\k';\,\chi_1,\chi_2,\hat{\n}_1,\hat{\n}_2 \right),
\ee
where the lightcone coordinates are functions of $\s$ and $\d$ via $\x_{1,2} = \d \pm \s/2$. (For simplicity we assume a mid-point definition of $\d$ in this schematic Appendix.)
A direct Taylor expansion of $\Pobs$ in powers of $\s$ would fail to converge because the pair separation is governed by two different dimensionless parameters: the Fourier phase $ks \gtrsim 1$ and the opening angle $s/d \ll 1$. Physically, the phase $e^{i\k'\cdot\s}$ in $\Pobs$ is due to the spatial translation between the points after averaging out observer positions. By extracting this dominant oscillatory component we can use the de-phased observer spectrum, 
\be\label{eq:F_def}
  \mathcal{F}(\k';\s,\d) \equiv e^{-i\k'\cdot\s}\; \Pobs(\k';\chi_1,\chi_2,\hat{\n}_1,\hat{\n}_2)\,,
\ee
the remaining $\s$-dependence is from the distortion of the lightcone geometry through the opening angle and radial evolution. For local observables, this  removes all $ks$ oscillations, leaving a function that varies slowly like $\mathcal{O}(s/d)$. For integrated observables, there are residual phase oscillations remain along the line of sight (e.g., contributions such as $e^{i k \mu_i (\chi'-\chi_i)}$ for lensing), but they are bounded by the projection geometry. In both cases,  the final  wide-angle series is in powers of $s/d$.

Because the underlying physical fields are real, exchanging the two legs ($\s \to -\s$) conjugates the observer spectrum: $\Pobs(\k';2,1) =[\Pobs(\k';1,2)]^*$. This implies the Hermitian symmetry:
\be\label{eq:F_hermitian}
  \mathcal{F}(\k';-\s,\d) = \mathcal{F}^*(\k';\s,\d)\,.
\ee
Consequently, $\text{Re}[\mathcal{F}]$ is an even function of $\s$, and $\text{Im}[\mathcal{F}]$ is an odd function of $\s$. At $\s=0$, the imaginary part vanishes, and the plane-parallel value $\mathcal{F}^{\rm PP} \equiv \mathcal{F}|_{\s=0}$ is real. This simplifies the wide angle expansion.

\subsubsection{The intrinsic wide-angle expansion}

We can parameterise the departure from the plane-parallel configuration in terms of the opening angle and radial separation, given by  $\delta\alpha^A = (\delta\mu,\,\delta\chi)$, where $\delta\mu = \mu_1 - \mu_2$ and $\delta\chi = \chi_1 - \chi_2$. These are physical observables (for fixed~$\k$) for any separation geometry in contrast to~$\s$ which is a derived vector lying `inside' the past lightcone.  Both~$\delta\alpha^A$ vanish at~$\s = 0$ and flip sign under~$\s\to-\s$.
Because of the Hermitian symmetry from Eq.~\eqref{eq:F_hermitian}, the Taylor expansion of $\mathcal{F}$ takes the form:
\be\label{eq:F_expand}
  \mathcal{F} = \mathcal{F}^{\rm PP} + i\,\delta\alpha^A\,\mathcal{D}_A + \frac{1}{2}\,\delta\alpha^A\,\delta\alpha^B\,\mathcal{W}_{AB} + \cdots\,,
\ee
where the {intrinsic scalars} are defined as:
\be\label{eq:DA_WAB_def}
  \mathcal{D}_A \equiv -i\, \frac{\partial \mathcal{F}}{\partial(\delta\alpha^A)} \bigg|_{\rm PP}\,, \qquad
  \mathcal{W}_{AB} \equiv \frac{\partial^2 \mathcal{F}}{\partial(\delta\alpha^A)\,\partial(\delta\alpha^B)} \bigg|_{\rm PP}\,,
\ee
where PP is the plane parallel limit and
the derivatives are evaluated at the fixed midpoint $\d$. The factor of $-i$ in $\mathcal{D}_A$ just gives a real quantity from the derivative of an odd imaginary function. The matrix $\mathcal{W}_{AB}$ is real and symmetric. 
These scalars isolate the intrinsic variation of the field statistics caused by the angular and radial opening of the lightcone legs. The separation between observers $\y$ does not appear here because it has already been Fourier-transformed into the wavevector $\k$.

\subsubsection{Conversion to the mixed spectrum}

We have to convert the physical deviations $\delta\alpha^A$ to the separation vector $\s$ to perform the FT to give $\bar P$. This is  via a  Jacobian, $\delta\alpha^A = J^A_i s_i + \mathcal{O}(s^2)$, evaluated at $\s=0$:
\be\label{eq:jacobian}
  \bm{J}^\mu = \frac{\hat{\bm{k}}' - \mu\,\hat{\bm{d}}}{d}\,, \qquad \bm{J}^\chi = \hat{\bm{d}}\,.
\ee
Substituting the expansion \eqref{eq:F_expand} into \eqref{eq:Pkd_def_app}, each factor of $s_i$ inside the Fourier transform becomes a derivative $-i\partial/\partial k'_i$ acting on the Dirac delta function $\delta^{(3)}_{\rm D}(\k'-\k)$. Integrating by parts moves these $\k'$-derivatives onto $\k$-derivatives acting on the intrinsic scalars and the Jacobians. This results in:
\be\label{eq:Pkd_result}
  \bar{P}(\k,\d) = \mathcal{F}^{\rm PP} - \nabla_{\!\k}\cdot \Big[\mathcal{D}_A\,\bm{J}^A\Big] - \frac{1}{2}\, \nabla_{\!\k}\!\cdot\!\Big(\nabla_{\!\k}\cdot \Big[\mathcal{W}_{AB}\, \bm{J}^A\!\otimes\!\bm{J}^B\Big]\Big) + \cdots
\ee
This  isolates the three distinct components of a wide-angle correction: the intrinsic field expansion in terms of the separation variables $\delta\alpha^A$ ($\mathcal{D}_A, \mathcal{W}_{AB}$), the lightcone geometry ($\bm{J}^A$), and the $\k$-space divergence ($\nabla_{\!\k}$) generated by the Fourier projection into pair separations.

\subsubsection{Application to local and integrated observables}

For local fields, the intrinsic scalars are polynomials in $\mu$, making the $\k$-space divergences straightforward to compute. For standard {Density + Kaiser RSD}, the parity-odd scalar vanishes ($\mathcal{D}_A = 0$) and the leading wide-angle effect is  second-order, governed by $\mathcal{W}_{\mu\mu}^{\rm RSD} = f(b - f\mu^2)D^2 P_0(k)$. Applying the double divergence in Eq.~\eqref{eq:Pkd_result} reproduces the standard Newtonian wide-angle corrections~\citep{1998ApJ...498L...1S,2000ApJ...535....1M,2004ApJ...614...51S,2008MNRAS.389..292P,2016JCAP...01..048R,2018MNRAS.476.4403C}. For {Doppler} effects, the odd-parity imaginary kernel gives $\mathcal{D}_A$ at first order, with $\mathcal{D}_\mu^{\rm Dopp} = \frac{f\mathcal{H}}{k}(b + f\mu^2)D^2 P_0(k)$. The single divergence $\nabla_{\!\k} \cdot [\mathcal{D}_A \bm{J}^A]$ gives the standard Doppler wide-angle monopole and dipole corrections~\citep{2023JCAP...04..067P,2024JCAP...08..027J,Noorikuhani:2022bwc}.

For {integrated observables} such as lensing, the de-phased spectrum takes the form $\mathcal{F}_{\rm Lens} \propto \int \diff\chi' W_L e^{ik\mu_1(\chi' - \chi_1)} \dots$. Taking the derivative to compute $\mathcal{D}_\mu$ brings a  factor of $ik(\chi' - \chi_1)$ down inside the line-of-sight integral. Because this factor is acted upon by the subsequent $\k$-space divergence in Eq.~\eqref{eq:Pkd_result}, the resulting wide-angle expansion remains governed by the parameter $\mathcal{O}(s/d)$ and can be systematically evaluated \citep[see also, e.g.,][to model multi-tracer integrated effects]{2026JCAP...06..039A,Noorikuhani:2022bwc,Guedezounme:2024pbj}.

%\clearpage 
\bibliographystyle{mnras}
\bibliography{ref}

\end{document}